\documentclass[pre,aps,twocolumn,superscriptaddress]{revtex4-2}

\usepackage{graphicx}
\usepackage{amssymb,amsfonts,amsmath}
\usepackage{color}
\usepackage{ulem}
\usepackage[hidelinks]{hyperref}
\usepackage{bbm}
\usepackage{bbold}

\usepackage{tikz}
\usetikzlibrary{shapes}
\usetikzlibrary{patterns}
\usetikzlibrary{angles, quotes}

\newcommand{\rv}{{\mathbf r}}

\newcommand{\ev}{{\bf e}}

\newcommand{\Tr}{{\rm Tr}\,}

\newcommand{\pv}{{\bf p}}

\newcommand{\Fv}{{\bf F}}

\newcommand{\msphantom}[1]{$\ldots$}

\renewcommand{\vec}{\mathbf}

\newcommand{\eps}{{\boldsymbol \epsilon}}
\newcommand{\rhotwo}{\rho^{(2)}}

\newcommand{\unit}{\vec{e}}

\begin{document}


\title{Force density functional theory in- and out-of-equilibrium}

\author{Salom\'ee M. Tschopp}
\affiliation{Department of Physics, University of Fribourg,
  CH-1700 Fribourg, Switzerland}

\author{Florian Samm\"uller}
\affiliation{Theoretische Physik II, Physikalisches Institut, 
  Universit{\"a}t Bayreuth, D-95447 Bayreuth, Germany}

\author{Sophie Hermann}
\affiliation{Theoretische Physik II, Physikalisches Institut, 
  Universit{\"a}t Bayreuth, D-95447 Bayreuth, Germany}

\author{Matthias Schmidt}
\affiliation{Theoretische Physik II, Physikalisches Institut, 
  Universit{\"a}t Bayreuth, D-95447 Bayreuth, Germany}
\email{Matthias.Schmidt@uni-bayreuth.de}

\author{Joseph M.\ Brader}
\affiliation{Department of Physics, University of Fribourg,
  CH-1700 Fribourg, Switzerland}

\date{\today}

\begin{abstract}
When a fluid is subject to an external field, as is the case  
near an interface or under spatial confinement, then the density 
becomes spatially inhomogeneous. Although the one-body density provides 
much useful information, a higher level of resolution 
is provided by the two-body correlations. 
These give a statistical description of the internal microstructure 
of the fluid and enable calculation of the average interparticle 
force, which plays an essential role in determining both the equilibrium 
and dynamic properties of interacting fluids. 
We present a theoretical framework for the description of 
inhomogeneous (classical) many-body systems, based explicitly on 
the two-body correlation functions. 
By consideration of local Noether-invariance against spatial distortion 
of the system we demonstrate the fundamental status of the 
Yvon-Born-Green (YBG) equation as a local force-balance within the fluid. 
Using the inhomogeneous Ornstein-Zernike equation we show that 
the two-body correlations are density functionals 
and, thus, that the average interparticle force entering the YBG 
equation is also a functional of the one-body density. 
The force-based theory we develop  provides an alternative to 
standard density functional theory for the study of inhomogeneous 
systems both in- and out-of-equilibrium. 
We compare force-based density profiles to the results of the 
standard potential-based (dynamical) density functional theory. 
In equilibrium, we confirm both analytically and numerically that the standard approach yields profiles that are consistent with the compressibility pressure, whereas the force-density functional gives profiles consistent with the virial pressure. 
For both approaches we explicitly prove the hard-wall contact theorem 
that connects the value of the density profile at the hard-wall with the bulk pressure. 
The structure of the theory offers deep insights into the nature of correlation in dense and inhomogeneous systems.
\end{abstract}

\maketitle 

\section{Introduction}
The analysis of spatial inhomogeneity is a primary means to
characterize a wide range of self-organized and complex states of
matter \cite{hansen2013}. Representative examples of systems and
effects with inherent position-dependence cover a wide range of soft
matter \cite{evans2019physicsToday,nagel2017}, including hydrophobic
solvation in complex environments \cite{levesque2012jcp}, desorption
of water at short and long length scales \cite{jeanmairet2013jcp},
liquids at hydrophobic and hydrophilic substrates characterized by
wetting and drying surface phase diagrams
\cite{evans2019pnas,evans2015prl}, critical drying of liquids
\cite{evans2016prl}, solvent-mediated forces between nanonscopic
solutes \cite{chacko2017}, electrolyte aqueous solutions near a solid
surface \cite{martinjimenez2017natCom}, layering in liquids
\cite{hernandez-munoz2019}, the structure of liquid-vapor interfaces
\cite{muscatello2017,tschopp2020} and locally resolved density
fluctuations \cite{evans2019pnas,evans2015prl,evans2016prl,chacko2017,
  eckert2020auxiliaryFields}.

Obtaining a systematic understanding of the physics that emerges in 
such systems can be achieved by using microscopically resolved 
correlation functions. 
In particular the one-body density profile captures a broad spectrum 
of behaviours, from strong oscillations in dense liquids, where molecular 
packing effects dominate \cite{levesque2012jcp, martinjimenez2017natCom,
  hernandez-munoz2019, muscatello2017, eckert2020auxiliaryFields}, to
pronounced drying layers near hydrophobic substrates when approaching
bulk evaporation \cite{jeanmairet2013jcp,evans2019pnas,evans2015prl,
  evans2016prl, chacko2017}. 
Effects such as these can be induced by walls or other external 
influence, which typically is modelled by a position-dependent external 
potential $V_{\text{ext}}(\rv)$.  
The physical relationship of the external potential
with the density profile $\rho(\rv')$ is often viewed in a causal way,
such that a change in the external potential at some position $\rv$
will create a density response in the system \cite{hansen2013}. In
general this response will not only occur at the same position, but
also, mediated by the interparticle interactions, at positions $\rv'$
further away. Near a surface phase transition
\cite{evans2019pnas,evans2015prl} the associated length-scale can
become very large.

On a formal level, $V_{\text{ext}}(\rv)$ and $\rho(\rv')$ form a pair of
conjugate variables within the variational framework of classical
density functional theory (DFT)
\cite{evans1979,evans1992,evans2016,hansen2013}.  DFT is based on the
existence of a generating (free energy) functional. Its nontrivial
contribution, the intrinsic excess free energy functional, $F_{\text{exc}}[\rho]$, originates
from the interparticle interactions. 
Due to inherent coupling of
the degrees of freedom of the many-body system, exact expressions for this quantity do not exist except for rare
special cases.  
Approximations are thus required for most applications. 
Minimizing the grand potential functional, $\Omega[\rho]$, typically by
numerically solving the associated Euler-Lagrange (EL) equation, then gives
results for the spatial structure and the thermodynamics of the
inhomogeneous system under consideration. 
The EL equation can be viewed as a
condition of local chemical equilibrium throughout the system
\cite{evans1979}. Here the local chemical potential consists of three
physically distinct contributions: a trivial ideal gas term, an excess
(over ideal) term which arises from the interparticle
interactions and an external contribution. 

An analogous point of view, which at first sight seems to be based on 
quite different physical intuition, is that of a force
balance relationship. As an equilibrium system is on average at rest,
the total local force must vanish at each point in space. This is a
classical result obtained by Yvon \cite{yvon1935}, Born and Green
\cite{born1946} (YBG) and it forms an exact property (sum-rule). 
Within computer simulation methodology, working on the level of force
distributions has recently received a boost through the introduction
of smart sampling strategies. `Use the force' \cite{rotenberg2020}
constitutes a new paradigm for obtaining data with significantly
reduced statistical noise
\cite{borgis2013,delasheras2018forceSampling,purohit2019}, as compared
to direct sampling via simple counting of events.
Force distributions naturally generalize to nonequilibrium, 
where the equilibrium ensemble average is replaced by a dynamical 
average over the corresponding set of states,
e.g.\ for overdamped Brownian
dynamics~\cite{delasheras2018velocityGradient,delasheras2020fourForces}. 
For quantum
systems, the locally resolved force-balance relationship was recently
addressed for dynamical situations \cite{tarantino2021,tchenkoue2019}.
Furthermore two-body correlation functions are central to
the recently developed conditional probability DFT \cite{mccarty2020,pederson2022}.

On a fundamental level it is apparent that out-of-equilibrium, it is forces, rather than potentials, 
that play the central role in determining the particle motion. 
A dynamical theory based on potentials will clearly be incapable of treating nonconservative forces and can also be expected to break down whenever the microstructure of the 
system deviates strongly from that of equilibrium. 
These difficulties present a fundamental limitation to the 
usefulness of existing dynamical density functional 
(DDFT) approaches 
\cite{evans1979,marconi99,archer_evans} 
and have served to motivate development of 
the force-based power functional theory (PFT) \cite{schmidt2013pft,schmidt2021pft}. 
 
Focusing on equilibrium, the YBG derivation conventionally rests on
formally integrating the full $N$-body equilibrium distribution over $N\!-\!1$ 
spatial degrees of freedom 
\cite{yvon1935,born1946,hansen2013}. 
The one-body density is thus expressed in terms of an integral of
the two-body density. 
In contrast, DFT is closed on the one-body level (using the EL equation) and hence neither requires consideration of the two-body level, nor does it permit to systematically incorporate such information.
%
%
In this paper we present a theoretical density functional approach,  which accounts explicitly for the interparticle forces and enables calculation of the one-body density for inhomogeneous 
fluids both in- and out-of-equilibrium. 

\section{Roadmap}\label{roadmap}
In the following we give an overview to guide the reader through the main results and concepts
presented in this work.
We begin, in subsection \ref{noether}, by developing the 
Noether theorem for the invariance of the 
grand potential under spatial distortions, as 
characterized by a vector displacement field $\eps(\rv)$. 
Expressing the grand potential as a functional of 
$\eps(\rv)$ leads to the variational condition,
\begin{equation*}
\frac{\delta \Omega[\eps]}{\delta \eps(\rv)}\Big|_{\eps(\rv)=0}
=\,0,
\end{equation*}
which generates the following force-balance (YBG) relation 
\begin{multline*}
-k_B T \, \nabla_{\rv_1}\ln\big(\rho(\rv_1)\big) - \nabla_{\rv_1}V_{\text{ext}}(\rv_1) \\
- \int d \rv_2 \frac{\rhotwo(\rv_1, \rv_2)}{\rho(\rv_1)} \nabla_{\rv_1} \phi(|\rv_1 - \rv_2|) = 0,
\end{multline*}
where the subscripted position variables $\rv_1$ and $\rv_2$ play the
role of a fixed point in space, $\rv_1$, and a `field point' which
is integrated over, $\rv_2$.
The two-body density and pair interaction potential are 
indicated by $\rhotwo$ and $\phi$, respectively 
($k_B$ denotes the Boltzmann constant and $T$ is the absolute
temperature). 
Our variational derivation highlights the fundamental status of the YBG equation, which we then take as a starting point for a self-consistent approach to determining the one-body density 
and can be written without approximation in the following form
\begin{equation*}
\rho(\rv_1) = {\rm e}^{\beta \left(\mu -V_{\text{ext}}(\rv_1)\right) + c_{\text{f}}^{(1)}(\rv_1)},
\end{equation*}
where $\mu$ is the chemical potential, 
$\beta\!=\!(k_BT)^{-1}$ and
the contribution $- k_BT c_{\text{f}}^{(1)}(\rv_1)$, acts as an effective external field arising from interparticle interactions. 
In subsection \ref{forceDFT}, we introduce the concept that the two-body correlation functions are functionals of the 
one-body density and we then use this to reinterpret the 
YBG equation as a {\it closed} integral equation for the one-body density. 
This leads us to the definition 
\begin{equation*}
c_{\text f}^{(1)}(\rv_1) \equiv -\nabla_{\rv_1}^{-1} \cdot \int d \rv_2 \frac{\rhotwo(\rv_1, \rv_2;[\rho])}{\rho(\rv_1)} \nabla_{\rv_1} \beta\phi(|\rv_1 - \rv_2|)\,,
\end{equation*}
in which the interparticle forces appear explicitly 
via $\nabla_{\rv_1}\phi$. 
The integral operator 
$\nabla_{\rv_1}^{-1}$ is defined in the main text and the square brackets indicate a functional dependence. 
An essential feature of our approach is that we have a computationally feasible scheme to evaluate the 
inhomogeneous density functional $\rhotwo(\rv_1, \rv_2;[\rho])$, 
which then leads to a closed, self-consistent `force-DFT'. 

In subsection \ref{potentialDFT}, our force-based approach
is contrasted with the standard DFT methodology (referred to in this work as potential-DFT) 
in which the grand potential is expressed as a functional 
of the one-body density and satisfies the 
variational condition,
\begin{equation*}
\frac{\delta\Omega[\rho]}{\delta\rho(\rv)}
\Big|_{\rho(\rv)=\rho_0(\rv)}
=\,0,
\end{equation*}
where $\rho_0(\rv)$ is the equilibrium density profile (the 
subscript will be omitted in the following). 
This leads to the well-known EL equation 
\begin{equation*}
\ln\rho(\rv) - \beta\left(\mu - V_{\text{ext}}(\rv)\right) 
- c_{\text p}^{(1)}(\rv)=0,
\end{equation*}
which can be expressed in the following alternative form
\begin{equation*}
\rho(\rv) = {\rm e}^{\beta \left(\mu -V_{\text{ext}}(\rv)\right) + c_{\text{p}}^{(1)}(\rv)},
\end{equation*}
where the function $c^{(1)}_{\text p}$ is defined as a 
functional derivative of the excess Helmholtz free energy,
\begin{equation*}
c_{\text p}^{(1)}(\rv)=-\frac{\delta \beta F_{\text{exc}}[\rho]}{\delta\rho(\rv)}.
\end{equation*}
In contrast to the force-DFT, the potential-DFT involves only one-body functions. We thus require only a single vector position, $\rv$, as an 
independent variable and there is no need to employ additional subscripts. 
The EL equation is the potential-DFT analogue of the YBG equation arising from invariance with respect to spatial distortions. 
%
%
If the free energy and the two-body density functionals are known only approximately, then 
the two approaches to DFT will lead in general to different density profiles for a given external field. 
This allows for deep insight into the inner workings of DFT.
Subsection \ref{potentialDFT} is intended primarily for readers who are less familiar with the details of potential-DFT.

In potential-DFT the well-used contact theorem predicts that the contact density at a hard planar 
wall is equal to $\beta P^{\,\text{c}}$, where $P^{\,\text{c}}$ is the compressibility pressure, see Reference~\cite{hansen2013} for its definition. 
For the force-DFT we find that the equivalent result links the contact density to 
$\beta P^{\,\text{v}}$, where $P^{\,\text{v}}$ is the virial pressure. 
In subsections \ref{virial_contact} and \ref{compressibility_contact}, we prove the corresponding contact theorem for both approaches.
These sum-rules are exact and hold within any reasonable approximation scheme.
If the reader is prepared to accept these assertions without 
proof, then both of these subsections can be passed-over on a 
first reading of the manuscript. 

As mentioned previously, in order to implement the force-DFT, we require a feasible method to obtain the density functional 
$\rhotwo(\rv_1,\rv_2;[\rho])$. 
Therefore, from that point onwards we will rely on approximation schemes.  
In subsections \ref{HS-FMT} and \ref{implementation}, we recall the fundamental measure theory (FMT) for hard-spheres and provide information about the numerical implementation. 
The FMT generates an explicit expression for the two-body direct correlation function, 
$c^{(2)}(\rv_1,\rv_2;[\rho])$, as a functional of the one-body density.  
The two-body direct correlation function, now uniquely determined by the one-body density, can be used as input to 
the inhomogeneous Ornstein-Zernike (OZ) equation, 
\begin{align*}
h^{}(\rv_1,\rv_2) = c^{(2)}(\rv_1,\rv_2) \!+\! 
\int\! d\rv_3\, h^{}(\rv_1,\rv_3)\rho(\rv_3) c^{(2)}(\rv_3,\rv_2),
\end{align*}
which is then a linear integral equation for determination 
of the total correlation function, $h$. 
By self-consistent solution of the OZ equation we obtain $h$ for any given one-body density; $h$ is thus a density functional. 
It has been shown that the inhomogeneous OZ equation can be numerically solved to high accuracy both in planar and spherical geometry \cite{tschopp2020, tschopp2021,attard}. Using the relation
\begin{equation*}
\rho^{(2)}(\rv_1, \rv_2) = \rho(\rv_1) \rho(\rv_2) \bigg(h(\rv_1, \rv_2) +1\bigg)
\end{equation*} 
then gives a clear self-consistent scheme 
to determine $\rho^{(2)}$ as a functional of the one-body density.
In subsection \ref{hardwall_numerics}, we show numerical results for the equilibrium density of 
hard-spheres at a hard-wall using both force- and potential-DFT, and validate the 
analytical predictions for the contact density. 
This demonstrates explicitly that the presented framework is not merely formal, but that it forms a concrete numerical scheme for the systematic study of inhomogeneous fluids.

In section \ref{dynamics}, we consider nonequilibrium 
systems subject to overdamped Brownian dynamics and show how the 
force-DFT allows calculation of the 
time-dependent density, $\rho(\rv,t)$. 
We argue that the force-DFT provides the most natural starting point for 
the development of a dynamical theory for the density, 
as it is the forces which are responsible for moving the particles.
Conservation of particle number dictates that 
the density obeys the continuity equation 
\begin{equation*}
\frac{\partial \rho(\vec{r}_1,t)}{\partial t} = - \nabla_{\vec{r}_1} \cdot \vec{j}(\vec{r}_1,t),
\end{equation*}
where $\vec{j}(\vec{r}_1,t)$ is the current, which needs to be specified to have a closed theory. 

In subsection \ref{forceDDFT}, we describe the force-DDFT, 
which is based on the following exact expression for the current
\begin{multline*}
\vec{j}(\vec{r}_1,t) \!=\! -D_0 \, \rho(\vec{r}_1,t) \Bigg( \nabla_{\vec{r}_1} \ln(\rho(\vec{r}_1,t))
+ \nabla_{\vec{r}_1} \beta V_{\text{ext}}(\vec{r}_1) \\
+ \int d \vec{r}_2 \, \frac{\rho^{(2)}(\vec{r}_1,\vec{r}_2,t)}{\rho(\vec{r}_1,t)} \nabla_{\vec{r}_1} 
\beta \phi(|\rv_1-\rv_2|) \Bigg),
\end{multline*}
where $D_0$ is the diffusion coefficient.
Using the previously described equilibrium functional for 
$\rhotwo$ yields 
a closed adiabatic theory for the one-body density. 
At each time-step the integral term is explicitly evaluated 
to obtain the average force due to interparticle interactions. 
In contrast, the familiar potential-DDFT, recalled in subsection \ref{potentialDDFT}, employs only one-body functions.  
The current in this case is given by 
\begin{multline*}
\!\!\!\!\vec{j}(\vec{r},t) \!=\! -D_0 \, \rho(\vec{r},t)
\nabla_{\vec{r}}\Bigg( \ln(\rho(\vec{r},t))
+ \beta V_{\text{ext}}(\vec{r})
-  c^{(1)}_{\text{p}}(\vec{r},t) \Bigg). 
\end{multline*}
In subsection \ref{numericaltrap}, we employ the FMT to 
generate numerical results for the density relaxation in a harmonic-trap and we compare the predictions of the force-DDFT with those of the potential-DDFT. 
This demonstrates that our force-based theory provides a firm basis for developing a systematic understanding of nonequilibrium phenomena. 
Finally, in section \ref{conclusions}, we draw our conclusions and give an outlook for future work.

\section{Equilibrium theory}
\subsection{Force-balance generated by Noether's theorem}\label{noether}

We begin our development of force-DFT by starting with the microscopic Hamiltonian 
and using invariance arguments.
Let us consider a classical system of $N$-particles described by position coordinates
$\rv_1,\ldots,\rv_N\equiv \rv^N$ and momenta $\pv_1,\ldots,\pv_N\equiv
\pv^N$. 
The Hamiltonian $H$ has the standard form consisting of kinetic,
internal and external potential energy contributions according to
\begin{align}
  H &= \sum_{i=1}^N \frac{\pv_i^2}{2m}
  + U_N(\rv^N) + \sum_{i=1}^N V_{\text{ext}}(\rv_i).
  \label{EQhamiltonianFDFT}
\end{align}
Here $m$ indicates the particle mass, $U_N$ denotes the
total interparticle interaction potential and $V_{\text{ext}}$
is an external one-body field.

We consider a canonical transformation on phase-space, parameterized
by a vector field $\eps(\rv)$ that describes a spatial displacement (`distortion') at
position $\rv$.  The transformation affects both coordinates and
momenta and is given by
\begin{align}
  \rv_i &\to \rv_i+\eps(\rv_i) \equiv \rv'_i,
  \label{EQcanonicalTransformationCoordinates}\\
    \pv_i &\to \pv_i- \nabla_{\rv_i}\eps(\rv_i) \cdot \pv_i
    \equiv \pv'_i,
  \label{EQcanonicalTransformationMomenta}
\end{align}
where the primes indicate the new phase-space variables and $\nabla_{\rv_i}$ denotes differentiation with respect to ${\rv_i}$.  We consider
the displacement field $\eps(\rv)$ and its gradient to be small.

The change in phase space variables affects the Hamiltonian and
renders it functionally dependent on the displacement field, $H\!\to\! H[\eps]$. 
Inserting transformations \eqref{EQcanonicalTransformationCoordinates}
and \eqref{EQcanonicalTransformationMomenta} into
 equation \eqref{EQhamiltonianFDFT} and expanding in the displacement field to
linear order yields
\begin{align}
  H[\eps] &= H_0 -
  \sum_{i=1}^N \frac{\pv_i\pv_i}{m}:\nabla_{\rv_i}\eps(\rv_i)
  \notag\\&\quad
  +\sum_{i=1}^N \eps(\rv_i)\cdot\nabla_{\rv_i}
  \big(U_N(\rv^N)+V_{\text{ext}}(\rv_i)\big),
  \label{EQhamiltonianTransformedFDFT}
\end{align}
where $H_0\!=\!H[\eps\!=\!0]$ is the original Hamiltonian 
as given in equation \eqref{EQhamiltonianFDFT}.
The colon indicates the contraction 
$\pv_i\pv_i\!:\!\nabla_{\rv_i}\eps(\rv_i)\!=\!
\sum_{\alpha,\gamma}p_{i\alpha}p_{i\gamma}\nabla_{r_{i\gamma}}
\epsilon_{\alpha}$, where Greek indices indicate Cartesian components.

Turning to a statistical description, the grand potential, 
$\Omega_0\!=\!\Omega[\eps\!=\!0]$, and
the grand partition sum, $\Xi_0\!=\!\Xi[\eps\!=\!0]$, of the original system are given respectively by
\begin{align}
  \Omega_0 &= -k_BT \ln \Xi_0\,,
  \label{EQgrandPotentialFDFT}\\
  \Xi_0 &= \Tr {\rm e}^{-\beta (H_0 - \mu N)}.
  \label{EQpartitionSumFDFT}
\end{align}
%
In the grand canonical ensemble the trace is defined as $\Tr \!=\!
\sum_{N=0}^\infty (h^{3N}N!)^{-1} \int d\rv_1\ldots d\rv_N
d\pv_1\ldots d\pv_N$, with $h$ indicating the Planck constant 
\cite{hansen2013}.

The transformed Hamiltonian, $H[\eps]$, can be used to define a correspondingly transformed grand potential functional, $\Omega[\eps]\!=\! -k_BT \ln \left(\Tr {\rm
  e}^{-\beta(H[\eps]-\mu N)}\right)$. To linear order in $\eps(\rv)$ the
functional Taylor expansion of $\Omega[\eps]$ is given by
\begin{align}
  \Omega[\eps] = \; &\Omega_0
  +\int d\rv\,
 \frac{\delta \Omega[\eps]}{\delta\eps(\rv)}
 \Big|_{\eps(\rv)= 0}\!\!\cdot\eps(\rv).
 \label{EQomegaTaylorExpansionFDFT}
\end{align}
The functional derivative in \eqref{EQomegaTaylorExpansionFDFT} can be calculated as follows
\begin{align}
  \frac{\delta \Omega[\eps]}{\delta\eps(\rv)}
  &=
  -\frac{k_BT}{\Xi[\eps]}\Tr
  \frac{\delta}{\delta\eps(\rv)}
       {\rm e}^{-\beta(H[\eps]-\mu N)}
  \notag\\&=
  -\frac{k_BT}{\Xi[\eps]}\Tr {\rm e}^{-\beta(H[\eps]-\mu N)}
  \Bigg(\!-\beta\frac{\delta H[\eps]}{\delta \eps(\rv)}\Bigg)
  \notag\\&=
  \Tr \Psi \frac{\delta H[\eps]}{\delta\eps(\rv)},
  \label{EQdeltaOmegaAsAverage}
\end{align}
where we have identified $\Psi\!=\!{\rm
  e}^{-\beta(H[\eps]-\mu N)}/\Xi[\eps]$ as the grand ensemble probability
distribution. Notably the form \eqref{EQdeltaOmegaAsAverage}
constitutes a grand ensemble average of $\delta H[\eps]/\delta
\eps(\rv)$.  
Formally, the average is taken in the displaced system, 
but we will find 
the form \eqref{EQdeltaOmegaAsAverage} to be sufficient to
calculate averages with respect to the original, undisplaced 
distribution. 
Using equation \eqref{EQhamiltonianTransformedFDFT} and thus retaining only the lowest relevant
order in $\eps(\rv)$ we obtain
\begin{align}\label{EQdeltaHdeltaEps}
  \frac{\delta H[\eps]}{\delta \eps(\rv)} &=
  \sum_{i=1}^N 
\Big(  
  -\frac{\pv_i\pv_i}{m} \cdot \nabla_{\rv_i} \delta(\rv-\rv_i)
  \\ &\qquad\qquad
+ \delta(\rv-\rv_i) \nabla_{\rv_i}
\big(
 U_N(\rv^N) + V_{\text{ext}}(\rv_i)
\big)
\Big).
\notag 
\end{align}
Here we have used the fundamental rule of functional differentiation
$\delta \eps(\rv)/\delta \eps(\rv')\!=\!\delta(\rv-\rv') {\mathbbm 1}$,
where $\delta(\cdot)$ indicates the (three-dimensional) Dirac
distribution, and ${\mathbbm 1}$ denotes the $3\!\times\! 3$ unit matrix.
Using equation \eqref{EQdeltaHdeltaEps} inside of the average
\eqref{EQdeltaOmegaAsAverage}, 
carrying out the phase-space
integrals, evaluating at $\rm \eps(\rv)= 0$ and multiplying by $-1$ yields
\begin{align}
  -\frac{\delta \Omega[\eps]}{\delta \eps(\rv)}
  \Big|_{\rm \eps(\rv)= 0}
  \!\!=-k_BT \nabla_{\rv} \rho(\rv)
  +\Fv_{\text{int}}(\rv) 
  - \rho(\rv) \nabla_{\rv} V_{\text{ext}}(\rv),
  \label{EQdeltaOmegaAsForcesFDFT}
\end{align}
where the one-body density profile is defined as the average
$\rho(\rv)\!=\!\Tr \Psi \sum_i\delta(\rv-\rv_i)$ and the internal force
density is given by $\Fv_{\text{int}}(\rv)\!=\!-\Tr \Psi \sum_i
\delta(\rv-\rv_i)\nabla_{\rv_i} U_N(\rv^N)$.  Furthermore the ideal
diffusion force density $-k_BT\nabla_{\rv}\rho(\rv)$ follows from carrying
out the phase-space momentum integrals explicitly or, alternatively,
using the equipartition theorem $\Tr\Psi \pv_i \partial H/\partial
\pv_i\!=\!k_BT \,{\mathbbm 1}$. The spatial gradients in
\eqref{EQdeltaOmegaAsForcesFDFT} emerge by 
exploiting $\nabla_{\rv}\,\delta(\rv-\rv_i)\!=\!-\nabla_{\rv_i}\delta(\rv-\rv_i)$. 
The right-hand side of
\eqref{EQdeltaOmegaAsForcesFDFT} represents the sum of the
position-resolved average one-body force densities of ideal, 
interparticle and external origin.

Using the transformations 
\eqref{EQcanonicalTransformationCoordinates} and
\eqref{EQcanonicalTransformationMomenta} one can easily verify that the differential volume
elements for coordinates and momenta follow to linear
order in $\eps$ as
\begin{align}
d\rv_i &\to  (1+\nabla_{\rv_i}\cdot\eps(\rv_i))d\rv_i  \equiv d\rv'_i, \\
d\pv_i &\to  (1-\nabla_{\rv_i}\cdot\eps(\rv_i))d\pv_i \equiv d\pv'_i.
\end{align}
We thus see that for each particle $d\rv_id\pv_i=d\rv'_i d\pv'_i$ holds to
linear order in $\eps$ and, therefore, for the entire phase-space
$\Pi_{i=1}^N d\rv_id\pv_i \!=\! \Pi_{i=1}^Nd\rv'_i d\pv'_i$,
as befits a canonical transformation
(see Appendix \ref{canonical}).

As the transformation is also time-independent, the Hamiltonian
  is an invariant (see again Appendix \ref{canonical}). Then trivially
  the partition sum \eqref{EQpartitionSumFDFT} and the grand potential
  \eqref{EQgrandPotentialFDFT} are also invariants. It follows that
\begin{align}
  \Omega[\eps] &= \Omega_0.
  \label{EQomegaEpsEqualsOmega}
\end{align}
The linear term in the functional Taylor expansion \eqref{EQomegaTaylorExpansionFDFT} thus vanishes and it does
so irrespective of the form of $\eps(\rv)$. 
The functional derivative \eqref{EQdeltaOmegaAsForcesFDFT} itself must therefore vanish,
\begin{equation*}
\frac{\delta \Omega[\eps]}{\delta \eps(\rv)}\Big|_{\eps(\rv)=0}
=\,0,
\end{equation*}
from which we can conclude that
\begin{align}
  -k_BT \nabla_{\rv} \rho(\rv) + \Fv_{\text{int}}(\rv) 
  - \rho(\rv) \nabla_{\rv} V_{\text{ext}}(\rv) &= 0,
  \label{EQforceDensityBalance}
\end{align}
which is the known equilibrium force density relationship 
\cite{schmidt2021pft,hansen2013}. 

When considering systems interacting via a pair potential $\phi$, 
the internal potential energy has the form $U_N(\rv^N) \!=\! \sum_{i<j}
\phi(|\rv_i-\rv_j|)$. The internal force
density can then be written as
\begin{align}
  \Fv_{\text{int}}(\rv_1)=-\int d\rv_2
  \rho^{(2)}(\rv_1,\rv_2)\nabla_{\rv_1}\phi_{12},
  \label{EQforceIntegral}
\end{align}
where the two-body density is defined microscopically as $\rho^{(2)}(\rv_1,\rv_2)\!=\!\Tr \Psi
\sum'_{ij} \delta(\rv_1-\rv_i) \delta(\rv_2-\rv_j)$, with the prime on the summation indicating the omission of the terms with $i=j$,
and we indicate the pair interaction potential by the shorthand $\phi_{12}\!=\!\phi(|\rv_1-\rv_2|)$.
We have relabelled $\rv\!\to\!\rv_1$ to give clarity to 
equations involving two-body functions. 
Using the explicit
form \eqref{EQforceIntegral} in the force density relationship
\eqref{EQforceDensityBalance} and rearranging yields 
\begin{multline}\label{YBG}
-k_B T \nabla_{\rv_1}\big(\ln\rho(\rv_1)\big) - \nabla_{\rv_1}V_{\text{ext}}(\rv_1) \\
- \int d \rv_2 \frac{\rhotwo(\rv_1, \rv_2)}{\rho(\rv_1)} \nabla_{\rv_1} \phi_{12} = 0 ,
\end{multline}
which is the explicit form of the first member of the YBG hierarchy \cite{hansen2013}.
We have thus shown that equation \eqref{YBG} arises from a variational principle on the grand potential.
It is hence of no lesser status than the EL equation of 
potential-DFT, to be discussed in subsection 
\ref{potentialDFT}.

\subsection{Force-DFT}\label{forceDFT}

The YBG equation \eqref{YBG}, which has been derived using Noether invariance in the previous subsection, has the appealing feature that it explicitly contains the interparticle pair interaction, $\phi_{12}$. 
We thus take the YBG equation \eqref{YBG} as a fundamental starting point for describing the equilibrium state. 
The third term in equation \eqref{YBG}, which gives the mean interparticle interaction force at the point $\rv_1$, is not written as the gradient of a potential. 
However, as an equilibrium system is conservative by construction we can formally rewrite it as a potential force using the inverse of the gradient 
\begin{multline}\label{star}
\int d \rv_2 \frac{\rhotwo(\rv_1, \rv_2)}{\rho(\rv_1)} \nabla_{\rv_1} \phi_{12}
\\=
\nabla_{\rv_1} \bigg( \nabla_{\rv_1}^{-1} \cdot \int d \rv_2 \frac{\rhotwo(\rv_1, \rv_2)}{\rho(\rv_1)} \nabla_{\rv_1} \phi_{12} \bigg),
\end{multline}
where $\nabla_{\rv_1}^{-1} \!=\! \frac{1}{4\pi} \int d \rv_2 \frac{(\rv_1-\rv_2)}{|\rv_1-\rv_2|^3}$ is an integral operator 
(see e.g.~\cite{delasheras2018forceSampling,rotenberg2020}). 
We thus {\it define} a scalar one-body function 
$c_{\text f}^{(1)}$ according to
\begin{equation} \label{def c1}
c_{\text f}^{(1)}(\rv_1) \equiv -\nabla_{\rv_1}^{-1} \cdot \int d \rv_2 \frac{\rhotwo(\rv_1, \rv_2)}{\rho(\rv_1)} \nabla_{\rv_1} \beta\phi_{12}.
\end{equation}
Although we employ the notation usually reserved for the one-body direct correlation function, equation \eqref{def c1} originates here from a quite different, but arguably more intuitive and fundamental way of thinking.
We can thus re-express the YBG equation \eqref{YBG} in the following form
\begin{equation*}
\nabla_{\rv_1} \bigg( -k_B T \ln\rho(\rv_1) - V_{\text{ext}}(\rv_1) + k_B T c_{\text f}^{(1)}(\rv_1) \bigg) = 0.
\end{equation*}
Equilibrium implies that the term in parentheses is equal to a constant,
which leads to 
\begin{equation}\label{EL force}
\rho(\rv_1) = {\rm e}^{\beta \left(\mu -V_{\text{ext}}(\rv_1)\right) + c_{\text f}^{(1)}(\rv_1)}.
\end{equation}
In contrast to standard potential-DFT here the function $c_{\text f}^{(1)}$ is simply defined by equation \eqref{def c1} and is generated directly from an explicit integral over the pair interaction force.
Combining equations \eqref{EL force} and \eqref{def c1} yields   
\begin{align}\label{main_forceDFT}
\rho(\rv_1) = \exp&\Bigg(\beta \left(\mu-V_{\text{ext}}(\rv_1)\right) 
\\
&- \nabla_{\rv_1}^{-1} \cdot \int d \rv_2 \frac{\rhotwo(\rv_1, \rv_2;[\rho])}{\rho(\rv_1)} \nabla_{\rv_1} \beta\phi_{12}\Bigg),
\notag
\end{align}
which is the central equation of force-DFT. 
Given an explicit expression for the two-body density as a functional of the one-body density, 
$\rhotwo\!\equiv\!\rhotwo(\rv_1,\rv_2;[\rho])$, equation \eqref{main_forceDFT} enables calculation of $\rho(\rv_1)$ for 
any given external potential. 
While one can argue on formal grounds that the force integral and
    its nontrivial essence, the two-body density distribution, are
    one-body density functionals, our current treatment makes this
    formal dependence both analytically explicit and computationally
    tractable.

\subsection{Potential-DFT}\label{potentialDFT}

The standard implementation of DFT (referred to in this work as potential-DFT) is based on the grand 
potential density functional, $\Omega[\rho]$, given by
\begin{equation}
\Omega[\rho] = F_{\text{id}}[\rho]  + F_{\text{exc}}[\rho]  - \int d\rv (\mu-V_{\text{ext}}(\rv)) \rho(\rv),
\end{equation}
where $\beta F_{\text{id}}[\rho]\!=\!\int d\rv \rho(\rv)\big( \ln(\rho(\rv))-1\big)$ is the ideal gas contribution with the 
thermal wavelength set equal to unity. 
Variational minimization of the grand potential 
\begin{equation}
\frac{\delta \Omega[\rho]}{\delta \rho(\rv)} = 0,
\end{equation} 
generates the EL equation,
\begin{equation}\label{EL potential}
\rho(\rv) = {\rm e}^{\beta \left(\mu -V_{\text{ext}}(\rv)\right) + c^{(1)}_{\text{p}}(\rv)}.
\end{equation}
The function $c^{(1)}_{\text{p}}$ is defined to be the first functional derivative 
of the excess (over ideal) Helmholtz free energy with respect to the density \cite{evans1979,hansen2013},
\begin{equation} \label{c1 potential}
c_{\text p}^{(1)}(\rv)=-\frac{\delta \beta F_{\text{exc}}[\rho]}{\delta\rho(\rv)}. 
\end{equation}
This function $c^{(1)}_{\text{p}}$, which is now the familiar one-body direct correlation function, is the first member of a hierarchy of 
correlation functions 
generated by successive functional differentiation of $F_{\text{exc}}$ with respect to the density. 
For situations in which all quantities are known exactly 
the definition given in equation \eqref{c1 potential} is 
equivalent to that of equation \eqref{def c1}.
Even though the EL equation \eqref{EL potential} has the same structure as equation \eqref{EL force}, these 
are conceptually different and have distinct origins.
The potential-DFT is constructed using only one-body functions and the average interaction force 
is generated by taking the gradient of $c^{(1)}_{\text{p}}$. 
This should be contrasted with the force-DFT, which works on the two-body level, in which the interaction force 
is calculated by explicit spatial integration of the pair-interaction, see equation \eqref{star}.

If both the one-body direct correlation function and the two-body density are generated from the same, exact 
free energy functional, then both the potential- and force-DFT implementations will yield the same average 
interaction force and thus the same density profiles. 
This will not be the case when using an approximate free energy functional and differences can be expected.  
The special case of a hard-wall substrate enables the degree of consistency between these two routes to be examined analytically and 
this will be the focus of the following two subsections \ref{virial_contact} and \ref{compressibility_contact}
(which can be skipped if the reader is more interested in the numerical predictions of potential- 
and force-DFT, which are presented in subsection \ref{hardwall_numerics}).
Route-dependency will also turn out to be highly relevant for 
the dynamical versions of potential- and force-DFT, as shown later in section \ref{dynamics}.

\subsection{Virial contact theorem}\label{virial_contact}


Before we proceed to investigate the virial route version of the contact theorem, we recall the very general and well-known version of it \cite{hansen2013,lebowitz1960,lovett1991}, namely that $\rho_{\text{w}}\!=\! \beta P$, which relates the density of a fluid at a planar hard-wall, $\rho_{\text{w}}$, to the corresponding bulk pressure, $P$. 
This can be proven without the need to specify by which method the one-body density is obtained. 
For completeness we provide a general proof of this in Appendix \ref{general_contact}. 
Other general proofs of the contact theorem are based on the balance of forces \cite{hansen2013,lebowitz1960,lovett1991}, a linear displacement of the free energy \cite{lovett1991} and the connection between the pressure and the mean kinetic energy density \cite{lebowitz1960}. 
The contact theorem is satisfied within DFT for excess free energy functionals within the weighted density approximation \cite{vanSwoll1989,tarazona1984}, with FMT \cite{roth2010} being an important example.

There are several generalizations of the wall theorem, which include a version for higher-body densities \cite{sigert1966} and extensions to hard-walls with additional soft particle-wall interactions \cite{henderson1979,henderson1983} as well as to non-planar locally curved hard-walls \cite{blum1994,upton1998}, for which one can also get a local version of the contact theorem \cite{bier2018}. Another important generalization is the extension to ionic liquids \cite{blum1978, henderson1979, carnie1981, mallarino2015}, where an additional term proportional to the squared surface charge arises in the contact theorem. 

In the aforementioned derivations, the contact value of the density is only related to a general bulk pressure. Exceptions are the work of Lovett and Baus \cite{lovett1991}, where the authors identify the virial pressure and the study of Tarazona and Evans \cite{tarazona1984}, where the contact theorem for the Percus-Yevick and the hypernetted chain approximation were determined. 
Due to approximations within theoretical descriptions the pressures from different routes do not necessarily agree with each other.
When used in potential-DFT studies it is always implicitly assumed that the relevant pressure is that of 
the compressibility route \cite{vanSwoll1989}, which we prove is indeed the case in the next subsection. 
For the force-DFT 
we prove here first that the relevant bulk pressure is that of the virial route.
The ability to access these two routes for inhomogeneous systems offers both the possibility of new insight into the formal structure of DFT
and a useful tool for constructing approximate functionals.

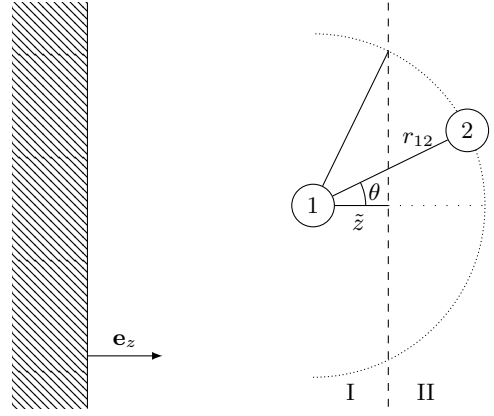
\begin{figure}
 \begin{minipage}[t]{0.4\textwidth}
 \hspace*{-0.5cm}
 \begin{tikzpicture}
 \coordinate (wall upper pt) at (-3,4.2);
 \coordinate (wall lower pt) at (-3,-1.2);
 \coordinate (wall size) at (-1,5.395);
 \coordinate (ez) at (-2.5,-0.5);
 \coordinate (z) at (0.6,1.5);
 \coordinate (rsph) at (1.4,2.175);
 \coordinate (I) at (0.5,-1.2);
 \coordinate (II) at (1.5,-1.2);
 \coordinate (dashed wall upper pt) at (1,4.2);
 \coordinate (dashed wall lower pt) at (1,-1.2);
 \coordinate (dashed wall upper cross) at (1,3.55);
 \coordinate (dashed wall lower cross) at (1,1.5);
 \coordinate (arc cross) at (2.28,1.5);
 \coordinate (origine arc) at (0,-0.785);
 \draw[densely dotted]  (origine arc) arc (-90:90:2.285); 
 \fill[white] (2.05,2.5) circle (8pt); 
 \node[draw, circle] (particle 1) at (0,1.5) {1};
 \node[draw, circle] (particle 2) at (2.05,2.5) {2};
 \draw (ez) node[above] {$\ev_{z}$};
 \draw (z) node[below] {$\tilde{z}$};
 \draw (rsph) node[above] {$r_{12}$};
 \draw (I) node[above] {I};
 \draw (II) node[above] {II};
 \draw[-, >=latex] (particle 1) -- (particle 2);
 \draw[->, >=latex] (-3,-0.5) -- (-2,-0.5);   
 \draw[-, >=latex] (particle 1) -- (dashed wall upper cross);
 \draw[-, >=latex] (particle 1) -- (dashed wall lower cross);
 \draw[loosely dotted, >=latex] (dashed wall lower cross) -- (arc cross);
 \draw[dashed, >=latex] (dashed wall upper pt) -- (dashed wall lower pt);
 \draw[-, >=latex] (wall lower pt) -- (wall upper pt);
 \fill[pattern=north west lines] (wall lower pt) rectangle ++(wall size);
 \pic [draw, -, angle radius=7mm, angle eccentricity=1.2, "$\theta$"] {angle = dashed wall lower cross--particle 1--particle 2};
 \end{tikzpicture}
 \end{minipage}
 \hspace*{2cm}
\caption{\textbf{Geometrical sketch of the planar geometry at a hard-wall.} For the evaluation of the virial integral 
(equations \eqref{eq:int} and \eqref{eq:z}) the space is divided into two sub-regions I and II. For a given value of coordinate 1 we integrate coordinate 2 over the angle $\theta\!=\!0\rightarrow \arccos(\tilde{z}/r_{12})$.
The shaded region on the left indicates the wall,
$\mathbf{e}_z$ denotes a unit vector in the $z$-direction 
and $\tilde{z}$ is the $z$-coordinate measured relative to 
coordinate $1$.
}
\label{sketch_geometry}
\end{figure}

Let us focus now on the virial contact theorem.
The force-DFT is generated by the YBG equation \eqref{YBG}. We begin by
spatially integrating it over the system volume $V$ to obtain
\begin{align}
 &- \int d\rv_1 \, \rho(\rv_1) \nabla_{\rv_1} \beta V_\text{ext}(\rv_1) \label{eq:ybg} \\
&\quad= \int d\rv_1 \,  \nabla_{\rv_1} \rho(\rv_1) + \int d\rv_1 \int d\rv_2 \, \rhotwo(\rv_1,\rv_2) \nabla_{\rv_1} \beta \phi_{12}. \nonumber
\end{align}
Exploiting the planar symmetry imposed by the hard-wall allows to simplify the density, $\rho(\rv_1)\!=\!\rho(z_1)$, the external potential 
$V_\text{ext}(\rv_1) \!=\! V_\text{ext}(z_1)$ and the spatial derivative $\nabla_{\rv_1} \!=\! \vec{e}_z \, d/dz_1 $, where $\vec{e}_z$ 
is the unit vector normal to the wall and pointing away from it (see Figure~\ref{sketch_geometry} for illustration).
The ideal contribution, i.e.\ the first term on the right-hand side of equation \eqref{eq:ybg}, can then be rewritten as 
\begin{align}
 \int d\rv_1 \, \nabla_{\rv_1} \rho(\rv_1)  =  A\!\! \int_{-\infty}^\infty d z_1 \, \frac{d \rho(z_1)}{d z_1} \vec{e}_z 
=   A \rho_{\text{b}} \vec{e}_z, \label{eq:id}
\end{align}
where $A \!=\! \int dx \int dy$ indicates the area of the hard-wall. 
In the second equality of \eqref{eq:id} we used that the density reaches a bulk value for large 
values of $z_1$, $ \rho(z_1 \!\to\! \infty)\!=\!\rho_{\text{b}}$, and vanishes inside the hard-wall, $\rho(z_1 \!\to\! - \infty)\!=\!0$. 
The external contribution, i.e.\ the left-hand side of equation \eqref{eq:ybg}, becomes
\begin{align}
- \!\! \int \!\!d\rv_1 \, \rho(\rv_1) \nabla_{\rv_1} \beta V_\text{ext}(\rv_1)&\!=\! 
-A\!\!\int_{-\infty}^\infty \!\!\!\!d z_1 \, \rho(z_1) \frac{d \beta V_\text{ext}(z_1)}{d z_1} \vec{e}_z 
\notag\\ 
&= A \rho_{\text{w}}\, \vec{e}_z. \label{eq:ext}
\end{align}
In deriving equation \eqref{eq:ext} we used that the derivative of the hard-wall external potential yields a (negative) delta distribution at the wall. 
To obtain this result it is useful to rewrite the density as 
$\rho(z_1)\!=\!n(z_1) \exp(-\beta V_\text{ext}(z_1))$, where $n(z_1)$ is a continuous function of $z_1$. 

The second term on the right-hand side of \eqref{eq:ybg} arises from the internal interparticle interactions and it is related to the global internal force, $\vec{F}_\text{int}^\text{\,o}$. Noether's theorem 
\cite{hermann2021noether} states that the global internal force vanishes in a closed system. 
As the semi-infinite system with a planar hard-wall is open, one has to take boundary contributions into account \cite{hermann2021noether}. The boundary terms corresponding to the $x$- and $y$-axis cancel due to the planar symmetry. 
The wall contribution, $z_1\!\to\! -\infty$, vanishes as there are no particles inside the hard-wall. 
(We refer the reader to Figure~\ref{sketch_geometry} to help visualize the situation for the following analysis.)

The remaining bulk boundary term, $z_1\!\to\!\infty$, can be treated by considering the force contributions 
between a particle inside a chosen integration volume and a particle outside of it. The force contributions 
where both particles are within the integration volume vanish for pair potential-type interparticle interactions, 
due to Newton's third law (\textit{actio equals reactio}). 
We choose the volume to be bounded from the right (positive values of $z$) by a virtual-plane parallel to the hard-wall and deep inside the bulk phase.
Therefore and because of the assumed finite interparticle interaction range the integrand 
and thus the two-body density, $\rhotwo(\rv_1,\rv_2)$, reduces to its bulk expression, $\rhotwo_{\text{b}}(|\rv_1-\rv_2|)$, at locations 
beyond this virtual separation plane. 
These considerations yield the following simplifications of the global internal force
\begin{align} 
- \beta \vec{F}_\text{int}^{\text{\,o}} &= \int d\rv_1 \int d \rv_2 \; \rhotwo(\rv_1,\rv_2) \nabla_{\rv_1} \beta \phi_{12}  
\notag\\
&= \int_\text{I} d \rv_1 \int_\text{II} d\rv_2 \; \rhotwo_{\text{b}}(r_{12}) \nabla_{\rv_1} \beta \phi_{12} 
\notag\\
&= \rho_{\text{b}}^2\int_\text{I} d \rv_1 \int_\text{II} d \rv_2 \; g(r_{12}) \frac{d \beta \phi_{12}}{d r_{12}} \cos\theta \vec{e}_z, 
\label{eq:int}
\end{align} 
where $r_{12}\!=\!|\rv_1-\rv_2|$. 
The subscript I on the integral denotes the integration volume with $z$-coordinate reaching from minus infinity to the volume boundary and II indicates the outside region at large $z$-values. 
To obtain the last relation in equation \eqref{eq:int} we use the identity 
$\rhotwo_{\text{b}}(r_{12}) \!=\! \rho_{\text{b}}^2 \, g(r_{12})$, where $g$ is the pair correlation function. 
The interparticle interaction force simplifies to 
$\nabla_{\rv_1} \phi(|\rv_1-\rv_2|)  \!=\! \vec{e}_z\cos(\theta)  d \phi(r_{12})/ d r_{12}$ due to the planar symmetry, 
where $\theta$ indicates the angle between the $z$-axis and the difference vector $\rv_1-\rv_2$.

We employ two different coordinate systems for each of the integration regions. 
For region I we use Cartesian coordinates, where $\tilde{z}$ measures the distance to the volume boundary.  
The integral over region II involves only the relative coordinate between the inside and outside regions. 
We express this integral in spherical coordinates, where the polar angle $\theta$ varies only between $0$ and $\tilde{\theta}\!=\!\arccos(\tilde{z}/r_{12})$ to ensure that the second coordinate 
remains within region II.
The $z$-component of equation \eqref{eq:int} is given by 
\begin{align}
&2 \pi A \rho_{\text{b}}^{2} \int_0^\infty \!\!\!dr_{12} \;  r_{12}^2 \,g(r_{12}) \frac{d \phi(r_{12})}{d r_{12}} \!
\int_0^{r_{12}} \!\!\!d\tilde{z}\! \int_{0}^{\tilde{\theta}}\!\!\! d \theta \, \sin\theta \cos\theta 
\notag\\
&= 2 \pi A \rho_{\text{b}}^2 \int_0^\infty \!\!\!d r_{12} \;  r_{12}^2 \,g(r_{12}) \frac{d \phi(r_{12})}{d r_{12}} \!\int_0^{r_{12}} \!\!\!d\tilde{z} \; \frac{1}{2}\left(1-\frac{\tilde{z}^2}{r_{12}^2} \right)  
\notag\\
&= \frac{2\pi}{3} A \rho_{\text{b}}^2 \int_0^\infty \!\!\!d r_{12} \; r_{12}^3 \,g(r_{12}) \frac{d \phi(r_{12})}{d r_{12}} . \label{eq:z}
\end{align}
Inserting equations \eqref{eq:id}, \eqref{eq:ext} and \eqref{eq:z} into the $z$-component of equation \eqref{eq:ybg} gives
\begin{align}\label{P virial}
\rho_{\text{w}}  &=  \rho_{\text{b}} - \frac{2 \pi}{3} \rho_{\text{b}}^2 \int_0^\infty d r_{12} \;  r_{12}^3 \,g(r_{12}) \frac{d \beta \phi(r_{12})}{d r_{12}} \notag\\
&= \beta P_{\text{id}} + \beta P_{\, \text{exc}}^{\, \text{v}} 
\quad= \beta P^{\, \text{v}},
\end{align}
where we have identified the standard expression \cite{hansen2013} for
the virial pressure $P^{\rm v}$.
We have thus proven the contact theorem relevant to the force-DFT, namely, that if one uses force-DFT to calculate the density profile at a hard-wall, then the contact density will correspond to the reduced virial pressure, $\beta P^{\, \text{v}}$.
Note that the derivation of the corresponding contact theorem in two dimensions can be done similarly.

\subsection{Compressibility contact theorem}\label{compressibility_contact}

The virial contact theorem derived above follows naturally from the forces acting within the system. 
In contrast the compressibility contact theorem, based on the one-body direct 
correlation function, is more formal and requires therefore more involved manipulations of the 
fundamental equations to arrive at the desired result.
Although the contact theorem is a result frequently cited in the literature, there is to our knowledge no calculation 
which shows explicitly that the wall contact density from potential-DFT is given by the reduced pressure from the \textit{compressibility} route. 
Since potential-DFT is generated by the EL equation \eqref{EL potential}, our proof begins by taking its gradient, which yields
\begin{multline}\label{comp_contact_thm_eq1}
\nabla_{\rv_1} \rho(\rv_1) + \rho(\rv_1) \nabla_{\rv_1} \beta V_{\text{ext}}(\rv_1) \\
-  \rho(\rv_1) \nabla_{\rv_1} c^{(1)}_{\text{p}}(\rv_1) = 0.
\end{multline}
The following simple identity 
from the product rule of differentiation
\begin{multline*}
\nabla_{\rv_1} \left(\rho(\rv_1) c^{(1)}_{\text{p}}(\rv_1)\right) \\
= c^{(1)}_{\text{p}}(\rv_1) \nabla_{\rv_1}\rho(\rv_1) + \rho(\rv_1) \nabla_{\rv_1} c^{(1)}_{\text{p}}(\rv_1), 
\end{multline*}
allows us then to rewrite \eqref{comp_contact_thm_eq1} in the following alternative form
\begin{multline}\label{comp_contact_thm_eq2}
\nabla_{\rv_1} \rho(\rv_1) + \rho(\rv_1) \nabla_{\rv_1} \beta V_{\text{ext}}(\rv_1) \\
-  \nabla_{\rv_1} \left(\rho(\rv_1) c^{(1)}_{\text{p}}(\rv_1)\right) + c^{(1)}_{\text{p}}(\rv_1) \nabla_{\rv_1}\rho(\rv_1) = 0.
\end{multline}
Equation \eqref{comp_contact_thm_eq2} involves only the one-body direct correlation function, $c^{(1)}_{\text{p}}$, defined in its standard form by equation \eqref{c1 potential}. 
However, 
the bulk compressibility pressure is typically expressed in terms of the two-body direct correlation function, $c^{(2)}$. 
Therefore, we seek to re-express $\rho(\rv_1) c^{(1)}_{\text{p}}(\rv_1)$ using the method of ``functional line integration''
\cite{line_integration}. 
By re-integrating the functional derivative of $\rho(\rv_1)c^{(1)}_{\text{p}}(\rv_1)$ with respect to the density we obtain 
the formal result
\begin{multline}
\rho(\rv_1) c^{(1)}_{\text{p}}(\rv_1) \\
= \int d\rv_2 \int_{0}^{\rho(\rv_2)} d\rho'(\rv_2) \left.\frac{\delta \Big(\rho(\rv_1) c^{(1)}_{\text{p}}(\rv_1)\Big)}{\delta \rho(\rv_2)}\right|_{\rho(\rv_2)=\rho'(\rv_2)},
\notag
\end{multline}
where at each spatial point $\rv_2$ we integrate from an empty system (zero density) up to the density of interest, 
$\rho(\rv_2)$.
Evaluation of the functional derivative then yields 
\begin{align}
\begin{split}
\rho(\rv_1) c^{(1)}_{\text{p}}(\rv_1) \!&=\!\! \begin{aligned}[t] \int d\rv_2 \int_{0}^{\rho(\rv_2)} \!\! d\rho'(\rv_2) \Big( 
\rho'(\rv_1) c^{(2)}(\rv_1, \rv_2 ; [\rho'])
\Big. \\ \Big.
 + \delta(\rv_1-\rv_2) c^{(1)}_{\text{p}}(\rv_1; [\rho'])  \Big)  \end{aligned} \notag \\
&=\!\!
\begin{aligned}[t]
\int d\rv_2 \int_{0}^{\rho(\rv_2)} \!\! d\rho'(\rv_2) \rho'(\rv_1) c^{(2)}(\rv_1, \rv_2 ; [\rho']) \\
+ \int_{0}^{\rho(\rv_1)} \!\! d\rho'(\rv_1) c^{(1)}_{\text{p}}(\rv_1; [\rho']). 
\end{aligned}
\end{split}
\end{align}
This result can be substituted into \eqref{comp_contact_thm_eq2} to give
\begin{align}\label{comp_contact_thm_eq3}
0 &= \nabla_{\rv_1} \rho(\rv_1) \;+\; \rho(\rv_1) \nabla_{\rv_1} \beta V_{\text{ext}}(\rv_1) \;+\; c^{(1)}_{\text{p}}(\rv_1) \nabla_{\rv_1}\rho(\rv_1) \notag\\
& \quad -  \int d\rv_2 \int_{0}^{\rho(\rv_2)} d\rho'(\rv_2) \nabla_{\rv_1} \Big(\rho'(\rv_1) c^{(2)}(\rv_1, \rv_2 ; [\rho'])\Big) \notag\\
& \quad -  \nabla_{\rv_1} \int_{0}^{\rho(\rv_1)} d\rho'(\rv_1) c^{(1)}_{\text{p}}(\rv_1; [\rho']).
\end{align}
We henceforth specialize to external fields which impose a planar geometry, such that the density profile and the two-body correlation functions 
exhibit cylindrical symmetry. 
As pointed out above, for the case of a planar hard-wall located at $z\!=\!0$ 
the density only varies in the $z$-direction, 
$\rho(\rv)\!=\!\rho(z)$. 
Equation \eqref{comp_contact_thm_eq3} can then be integrated to yield
\begin{multline}\label{comp_contact_thm_eq4}
\rho_{\text{b}} - \rho_{\text w} =\\
 \overbrace{ \int d\rv_2 \int_{0}^{\rho(z_2)} d\rho'(z_2) \rho'_{\text{b}} c^{(2)}(z_1\!=\!\infty, z_2, r_2; [\rho'])}^{\text{A}} 
\\
+\underbrace{ \int_{0}^{\rho_{\text{b}}} d\rho'_{\text{b}} c^{(1)}_{\text{p}, \text{b}}(\rho'_{\text{b}})}_{\text{B}} - \underbrace{\int_{-\infty}^{\infty} dz_1 c^{(1)}_{\text{p}}(z_1) \frac{d}{dz_1}\rho(z_1)}_{\text{C}},
\end{multline}
where we have used $ \rho(z_1 \!\to\! \infty)\!=\!\rho_{\text{b}}$ 
and $\rho(z_1 \!\to\! - \infty)\!=\!0$, as in equation \eqref{eq:id}.
To connect this expression with the bulk compressibility pressure we 
analyze each of the three terms labelled A, B and C in equation \eqref{comp_contact_thm_eq4} separately.

\underline{Term A}:
Having $z_1\!\rightarrow\!\infty$ as an argument of the two-body correlation function $c^{(2)}$ (which is of finite range) 
has the consequence that only bulk values contribute to the integral over the coordinate labelled $2$, thus
\begin{align*}
\quad \text{Term A} &=  \int_{0}^{\rho_{\text{b}}} d\rho'_{\text{b}} \, \rho'_{\text{b}} \int d\rv_{12} \, c^{(2)}_{\text{b}}(r_{12}; [\rho'_{\text{b}}]) \\
&=  \int_{0}^{\rho_{\text{b}}} d\rho'_{\text{b}} \, \rho'_{\text{b}} \tilde{c}^{(2)}_{\text{b}}(q\!=\!0; \rho'_{\text{b}}) \\
&= - \beta P_{\, \text{exc}}^{\, \text{c}},
\end{align*}
where $\tilde{c}^{(2)}_{\text{b}}(q\!=\!0)$ is the Fourier transform of the two-body direct correlation function in the zero wavevector limit.
The second equality gives the well-known integral giving the excess (over ideal) pressure in the compressibility route, 
$P_{\text{exc}}^{\,\text{c}}$, see Reference \cite{luchko}.

\underline{Term B}: 
This term does not require further manipulation and it can be given a clear physical interpretation. 
By identifying 
the bulk one-body direct correlation function, $c^{(1)}_{\text{p}, \text{b}}$, with the excess reduced chemical potential, $\mu_{\text{exc}}$, 
it follows that
\begin{align*}
 \int_{0}^{\rho_{\text{b}}} d\rho'_{\text{b}} c^{(1)}_{\text{p}, \text{b}}(\rho'_{\text{b}}) &= -\int_{0}^{\rho_{\text{b}}} d\rho'_{\text{b}}\, \beta \mu_{\text{exc}}(\rho'_{\text{b}}) \\
&= - \int_{0}^{\rho_{\text{b}}} d\rho'_{\text{b}} \,\beta \frac{\partial f_{\text{exc}}}{\partial \rho'_{\text{b}}} \\
&= - \beta f_{\text{exc}}(\rho_{\text{b}}),
\end{align*}
where $f_{\text{exc}}\!=\!F_{\text{exc}}/V$ is the bulk excess Helmholtz free energy per unit volume.

\underline{Term C}: Using that $c^{(1)}_{\text{p}}$ evaluated at a bulk density becomes position independent,
\begin{align*}
- \int_{-\infty}^{\infty} dz_1  \frac{d \rho(z_1)}{dz_1} c^{(1)}_{\text{p}}(z_1) &= - \int_{0}^{\rho_{\text{b}}} d\rho'_{\text{b}} c^{(1)}_{\text{p}}(z_1; [\rho'_{\text{b}}]) \\
&= - \int_{0}^{\rho_{\text{b}}} d\rho'_{\text{b}} c^{(1)}_{\text{p},\text{b}}(\rho'_{\text{b}}) \\
&= \quad \,\, \text{Term B}.
\end{align*}

Terms B and C cancel out and we finally get
\begin{align}\label{P compressibility}
\rho_{\text w} &= \rho_{\text{b}} - \int_{0}^{\rho_{\text{b}}} d\rho'_{\text{b}} \, \rho'_{\text{b}} \tilde{c}^{(2)}_{\text{b}}(q\!=\!0; \rho'_{\text{b}}) 
\notag\\
&= \beta P_{\text{id}} + \beta P_{\, \text{exc}}^{\, \text{c}} 
\quad= \beta P^{\, \text{c}}.
\end{align}
We have thus proven the contact theorem for the compressibility route,  
namely, that if one uses equation \eqref{comp_contact_thm_eq1} to calculate the density profile 
at a hard-wall, then the contact density will correspond to the reduced compressibility pressure, $\beta P^{\, \text{c}}$.

So far all our definitions and analytical considerations were not constrained to any specific system. 
However, at this point, in order to implement these general frameworks and show numerical results, 
we will focus on a particular simple model.

\subsection{Hard-sphere FMT}\label{HS-FMT}

We now specialize to the minimal fluid model, which we take to be hard-spheres of radius $R$ in three dimensions. 
The force-DFT approach is in no way restricted to this particular system. 
Hard-spheres simply provide a convenient test-case for which FMT gives
an accurate approximation to the excess Helmholtz free energy functional,
\begin{align}\label{ros_fe}
\beta F_{\text{exc}}[\,\rho\,] = \int d\rv_1 \; \Phi \left( \left\lbrace n_{\alpha}(\rv_1) \right\rbrace  \right).
\end{align}
The original Rosenfeld formulation of FMT \cite{rosenfeld89} employs the following reduced excess free energy density 
\begin{equation*}
\Phi = - n_0 \ln(1-n_3) + \frac{n_1 n_2 - {\bf n}_1 \cdot {\bf n}_2}{1-n_3} + \frac{n_2^3 
- 3 n_2 {\bf n}_2 \cdot {\bf n}_2}{24 \pi (1-n_3)^2}.
\end{equation*}
The weighted densities are generated by convolution	
\begin{equation}\label{n_alpha_ros_fe}
n_{\alpha}(\rv_1) = \int d\rv_2 \; \rho(\rv_2)\, \omega_{\alpha}(\rv_1-\rv_2), 
\end{equation}
where the weight functions, $\omega_{\alpha}$, are characteristic of the geometry of the spheres. 
Of the six weight functions, four are scalars 
\begin{align}
\omega_3(\rv)&=\Theta(R-r), \hspace*{0.5cm}
\omega_2(\rv)=\delta(R-r), \notag\\
\omega_1(\rv)&=\frac{\delta(R-r)}{4\pi R}, \hspace*{0.51cm}
\omega_0(\rv)=\frac{\delta(R-r)}{4\pi R^2}, \notag
\end{align}
and two are vectors (indicated by bold indices)
\begin{align}
\omega_{\bold 2}(\rv)&=\unit_{\rv}\,\delta(R-r),\hspace*{0.5cm}
\omega_{\bold 1}(\rv)=\unit_{\rv}\frac{\delta(R-r)}{4\pi R},
\notag
\end{align}
where $\unit_{\rv}=\rv/r$ is a unit vector.

Applying the definition \eqref{c1 potential} 
for $c_{\text{p}}^{(1)}$ to the free energy \eqref{ros_fe} generates 
the following approximate form for the one-body direct correlation function 
\begin{align}\label{c_onebody}
c^{(1)}_{\text{p}}(\rv_1)=-\sum_{\alpha}\int d\rv_2\, 
\Phi'_{\alpha}(\rv_2)\,\omega_{\alpha}(\rv_{21}),
\end{align}
where the summation runs over all scalar and vector indices, 
$\Phi'_{\alpha}\!=\!\partial \Phi/\partial n_{\alpha}$, and 
$\rv_{21}=\rv_2-\rv_1$. 
The function $\Phi'_{\alpha}$ is a vector quantity when $\alpha$ takes the value ${\bold 1}$ or ${\bold 2}$, 
in which case a scalar product with the corresponding vectorial weight function is implied in equation \eqref{c_onebody}, otherwise it is a scalar function.

Taking two functional derivatives of the free energy \eqref{ros_fe} generates the following 
expression for the two-body direct correlation function
\begin{align}
c^{(2)}(\rv_1, \rv_2)
&=-\!\sum_{\alpha\beta}\int \!d\rv_3\, 
\omega_{\alpha}(\rv_{31})\,
\Phi''_{\alpha \beta}(\rv_3)\,
\omega_{\beta}(\rv_{32}),
\label{c_ros_fe}
\end{align}
where $\Phi''_{\alpha \beta}\!=\!\partial^2 \Phi/\partial n_{\alpha} \partial n_{\beta}$. 
For detailed descriptions of how to implement equation \eqref{c_ros_fe} in planar 
and spherical geometries we refer the reader to Reference \cite{tschopp2021}. 

The inhomogeneous OZ equation
\begin{align}\label{oz}
h^{}(\rv_1,\rv_2) = c^{(2)}(\rv_1,\rv_2) \!+\! 
\int\! d\rv_3\, h^{}(\rv_1,\rv_3)\rho(\rv_3) c^{(2)}(\rv_3,\rv_2),
\end{align}
connects the two-body direct correlation function, $c^{(2)}$, with the total correlation function, $h$.
The latter is related to the two-body density according to
\begin{equation}\label{rhotwo_h}
\rho^{(2)}(\rv_1, \rv_2) = \rho(\rv_1) \rho(\rv_2) \bigg(h(\rv_1, \rv_2) +1\bigg).
\end{equation}
Substitution of equation \eqref{c_ros_fe} into the inhomogeneous OZ equation \eqref{oz} yields a linear integral equation 
which can be solved for $h$, given the one-body density as input. 
The two-body density defined by equation \eqref{rhotwo_h} is thus a known functional 
of the one-body density, as required for implementation of the force-DFT.

Numerical evaluation of the right-hand side of equation \eqref{c_ros_fe} followed 
by iterative solution of the inhomogeneous OZ equation \eqref{oz} is a demanding, yet well-defined and ultimately manageable, task. 
For researchers familiar with standard potential-DFT implementations (which operate purely on the 
one-body level) working with two-body numerics represents a significant step. 
However, having explicit access to the two-body correlations provides a much deeper insight into the 
particle microstructure and this benefit thus outweighs the increased computational complexity.

\subsection{Implementation in planar geometry}\label{implementation}
 
Now that we have specified the model of interest (hard-spheres) we choose to henceforth 
restrict our attention to planar geometry,
for which the two-body correlation functions can be expressed using the cylindrical coordinates, $z_1$, $z_2$ and $r_2$ 
(see References \cite{tschopp2020,tschopp2021}). 
Although neither the potential- nor the force-DFT are limited to any particular geometry, our choice to focus on the planar case enables us 
to make connection to the contact sum-rules proven analytically in subsections \ref{virial_contact} and \ref{compressibility_contact}.

In order to implement force-DFT, as expressed by the central 
equation \eqref{main_forceDFT}, we begin by integrating equation \eqref{def c1} to obtain the function $c_{\text{f}}^{(1)}$. This yields 
\begin{align} \label{eq:cf_hardsphere}
&c^{(1)}_{\text f}(z) - c^{(1)}_{\text f}(0) \\
&\quad = \int_0^z dz_1 \frac{2\pi}{\rho(z_1)} \int_{-\infty}^{\infty} dz_2 (z_1-z_2) \rhotwo(z_1,z_2,r_2^{*}),\notag
\end{align}
with $r_2^{*} \!\equiv\! \sqrt{1-(z_1-z_2)^2}$ and where the particle diameter has been set to unity. 
Both the factor $(z_1\!-\!z_2)$ and the argument 
$r_2^*$ appearing in the two-body density are consequences of 
the gradient of the hard-sphere potential in equation 
\eqref{def c1}; a detailed derivation of equation \eqref{eq:cf_hardsphere} is given in Appendix~\ref{Appendix_cOneForce}.
The integration constant $c^{(1)}_{\text f}(0)$ is unknown, but it does not have 
to be determined in order to calculate the density profile.
Defining a new parameter $\alpha \!\equiv\! \beta \mu + c_{\text f}^{(1)}(0)$, we obtain the 
following expression
\begin{align}\label{alpha_route_eq1}
\rho(z) &= {\rm e}^{\beta \left(\mu - V_{\text{ext}}(z)\right) + c^{(1)}_{\text f}(z)} \\
&= {\rm e}^{\alpha} {\rm e}^{-\beta V_{\text{ext}}(z) + \int_0^z dz_1 \frac{2\pi}{\rho(z_1)} \int_{-\infty}^{\infty} dz_2 (z_1-z_2) \rhotwo(z_1,z_2,r_2^{*})}. \notag
\end{align}
If we choose the average number of particles $\langle N \rangle\!=\!\int_{-\infty}^{\infty} dz \,\rho(z)$ to be conserved 
then the corresponding value of $\alpha$ can be determined from
\begin{equation}\label{alpha_route_eq2}
{\rm e}^{\alpha} \!=\! \frac{\langle N \rangle}{\int_{-\infty}^{\infty} \!\! dz \, {\rm e}^{-\beta V_{\text{ext}}(z) + 2\pi \! \int_0^z dz_1 \! \int_{-\infty}^{\infty} \! dz_2 \, (z_1-z_2) \frac{\rhotwo(z_1,z_2,r_2^{*})}{\rho(z_1)}}}, 
\end{equation}
which circumvents the need to prescribe $c_{\text f}^{(1)}(0)$ and $\mu$ independently.
Given a method to calculate the two-body density from a given one-body profile, equations \eqref{alpha_route_eq1} and 
\eqref{alpha_route_eq2} provide a closed system for numerical determination of the equilibrium density profile.
This is possible since the two-body direct correlation function, $c^{(2)}$, is given as a functional of the one-body density in equation 
\eqref{c_ros_fe}. 
The connection between $c^{(2)}$ and $\rhotwo$ is given by combining the inhomogeneous OZ equation \eqref{oz} with the 
definition \eqref{rhotwo_h}. 
The three-dimensional integral appearing in \eqref{oz} can be reduced to a manageable one-dimensional integral using 
the method of Hankel transforms, as described in detail in Reference \cite{tschopp2021}.
The Hankel transform of the OZ equation \eqref{oz} is given by
\begin{align}\label{oz_planar}
\overline{h}(z_1,z_2,k) &= \overline{c}^{\,(2)}(z_1,z_2,k) 
\\
&+ \int_{-\infty}^{\infty} \!dz_3 \;  
\overline{h}(z_1,z_3,k)\,\rho(z_3)\,\overline{c}^{\,(2)}(z_3,z_2,k), 
\notag
\end{align}
where an overbar indicates a Hankel transformed quantity. 
In Reference \cite{tschopp2021} a convenient analytical expression is given for the Rosenfeld form of $\overline{c}^{\,(2)}$.

On the other hand, 
the implementation of potential-DFT to obtain the density profile in planar geometry is a standard procedure in 
FMT studies. 
The EL equation \eqref{EL potential} in planar geometry reads
\begin{equation}\label{EL_planar}
\rho(z) = {\rm e}^{\beta \left(\mu - V_{\text{ext}}(z)\right) + c^{(1)}_{\text p}(z)}, 
\end{equation}
where $c^{(1)}_{\text p}(z)$ is given by the planar version of equation \eqref{c_onebody} (see Reference \cite{tschopp2021}).

\subsection{Numerical results for hard-spheres at a hard-wall}\label{hardwall_numerics}

\begin{figure*}
\center{\includegraphics[width=0.95\linewidth]{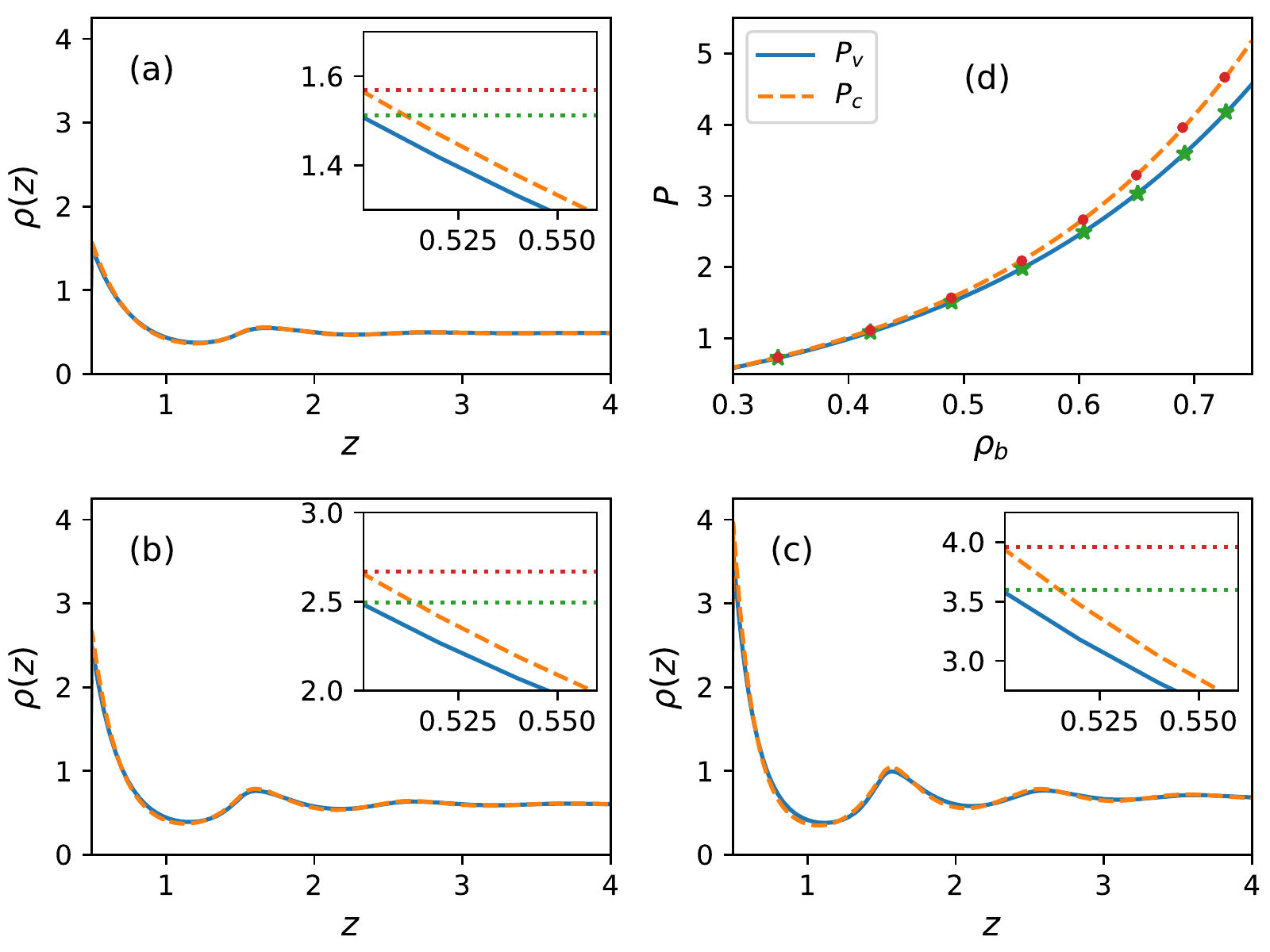}}
\caption{ {\bf Density profiles and pressure at a hard-wall.}
  Numerical results for hard-spheres at a hard-wall calculated using
  potential-DFT (dashed orange lines), force-DFT (full blue lines).  The density curves
  from potential-DFT shown in the
  panels (a), (b) and (c) were calculated at reduced chemical
  potentials $\beta\mu\!=\!3$, $5$ and $7$, respectively. The
  corresponding force-DFT density curves shown in the same
    panels were obtained by setting the average number of particles
  to match the values from potential-DFT.  The top right panel
    (d) shows the analytic PY compressibility and virial pressures,
  together with numerical contact-values from potential-DFT (red
  filled circles) and force-DFT (green stars).
  These are also indicated by dotted horizontal lines in the zoomed
  density profiles shown in the insets in panels (a), (b) and
    (c), following the same color scheme.    
\label{slit_fig}
}
\end{figure*}

To calculate the density profile from force-DFT, we need to choose as input a value for the average number of particles, $\langle N\rangle$.
In contrast, the potential-DFT takes as input the reduced chemical potential, $\beta \mu$.
To enable comparison of the results from the two different approaches, we
first calculate the potential-DFT density profiles for a given value of the reduced chemical potential
and then calculate the average number of particles in the system by spatial integration.
This value is then used as input in the force-DFT calculation.
The quantities of relevance for testing the wall contact theorem are the bulk density (taken as $\rho_{\text{b}}\approx\rho(z\!\to\!\infty)$) 
and the density at the wall, $\rho_{\text w}$. 
As we employ the Rosenfeld functional, the compressibility and virial pressures are identical to those of the 
well-known Percus-Yevick (PY) integral equation theory \cite{hansen2013}, which are given by
\begin{align}
\frac{\beta P^{\,\text{c}}}{\rho_{\text{b}}} &= \frac{1+\eta+\eta^2}{(1-\eta)^3}, 
\notag\\
\frac{\beta P^{\,\text{v}}}{\rho_{\text{b}}} &= \frac{1+2\eta+3\eta^2}{(1-\eta)^2},
\label{pressure}
\end{align}
where $\eta\!=\!\frac{4\pi}{3}\rho_{\text{b}} R^3$ is the packing fraction.

In Figure \ref{slit_fig}, we show three sets of representative density profiles and the contact density 
as a function of $\rho_{\text{b}}$ compared with the expected pressure. 
In the top left panel of Figure \ref{slit_fig}(a) the reduced chemical potential is rather low, $\beta\mu\!=\!3$, and the resulting potential- 
and force-DFT density profiles are very similar. 
Zooming to inspect the contact value highlights the difference between the two profiles and 
shows that our numerical data are highly consistent with 
the expected analytical pressures (indicated by the horizontal dotted lines, red for the compressibility 
route and green for the virial route). 
The bottom left panel of Figure \ref{slit_fig}(b) is for $\beta\mu\!=\!5$ and, although some slight differences begin to emerge in the oscillations 
of the two density profiles, the contact values remain in excellent agreement with the respective analytical predictions. 
The bottom right panel of Figure \ref{slit_fig}(c) is for $\beta\mu\!=\!7$ and shows more significant deviation of the density oscillations, but the contact densities still remain consistent with equations \eqref{pressure}. 
We find that both the oscillation amplitude and contact density from force-DFT are lower than those of the 
potential-DFT.

We have performed similar calculations for a wider set of reduced chemical potentials. 
The upper right panel of Figure \ref{slit_fig}(d) shows the resulting contact densities as a function of $\rho_{\text{b}}$.
This discrete set of points are shown together with the analytical curves from equations \eqref{pressure}, exhibiting 
an excellent level of agreement for a broad range of bulk densities. 
From our numerical results it is clear that the force-DFT does correspond to the virial route. 
This demonstrates that we have constructed a method by which DFT calculations can be reliably performed within the `virial realm' instead of the `compressibility realm', which seemed to be the only one accessible before.

\section{Dynamical theory}\label{dynamics}

\subsection{Force-DDFT}\label{forceDDFT}

The tools we have developed can be readily extended to explore the dynamics of the one-body density 
out-of-equilibrium. 
In the following we consider systems subject to overdamped Brownian dynamics (BD).
These model dynamics are suitable for the present investigation
  for two primary reasons. First, in overdamped BD the temperature is
  per construction constant. Hence relating the dynamics to an
  equilibrium ensemble is more straightforward than it is in Molecular
  Dynamics. Second, the absence of inertia in BD leads to simpler
  dynamical behaviour emerging on the one-body level
  \cite{schmidt2021pft}. An example is acceleration-dependent
  viscosity, which arises in Molecular Dynamics, but not in overdamped
  BD \cite{renner2022}.

For overdamped motion the dynamics of the $N$-body distribution function is dictated by the Smoluchowski equation 
\cite{archer_evans}. Integration over $N\!-\!1$ position coordinates generates the exact equation 
of motion for the one-body density
\begin{equation}\label{continuity}
\frac{\partial \rho(\vec{r}_1,t)}{\partial t} = - \nabla_{\vec{r}_1} \cdot \vec{j}(\vec{r}_1,t),
\end{equation}
where the current is given by 
\begin{multline}\label{exact_current}
\vec{j}(\vec{r}_1,t) = -D_0 \, \rho(\vec{r}_1,t) \Bigg( \nabla_{\vec{r}_1} \ln(\rho(\vec{r}_1,t))
+ \nabla_{\vec{r}_1} \beta V_{\text{ext}}(\vec{r}_1) \\
+ \int d \vec{r}_2 \, \frac{\rho^{(2)}(\vec{r}_1,\vec{r}_2,t)}{\rho(\vec{r}_1,t)} \nabla_{\vec{r}_1} \beta \phi_{12} \Bigg),
\end{multline}
where $D_0$ is the diffusion coefficient. 
The interparticle force, $-\nabla_{\vec{r}_1} \beta \phi_{12}$, appears explicitly in the integral term.
Equations \eqref{continuity} and \eqref{exact_current} form the basis of the force-DDFT. 
Calculation of the current requires the exact time-dependent two-body density, $\rhotwo$, as an input quantity, which is not available for any interacting model of real interest. 
A workable approximation can be obtained by making the assumption that $\rhotwo$ is instantaneously equilibrated to the nonequilibrium density. 
This adiabatic approximation enables one to employ the two-body 
correlations calculated using the inhomogeneous OZ equation \eqref{oz} (which is an equilibrium relation) to obtain the average interaction force at each time-step.
Note that in equilibrium the current \eqref{exact_current} vanishes. 
Since the density is nonzero, the sum of the three terms in parentheses in equation \eqref{exact_current} must 
also vanish and we recover the YBG equation \eqref{YBG}. 
The time-dependent density of force-DDFT thus relaxes to the density profile of force-DFT in 
the long-time limit.  

As for the equilibrium case, we only consider hard-spheres subject to external fields of planar geometry. 
The gradient inside the integral term of equation \eqref{exact_current} must therefore be treated 
carefully to correctly capture the discontinuous hard-sphere interaction potential. 
Fortunately, for planar geometry the integral can be conveniently reduced to one-dimension and 
equations \eqref{continuity} and \eqref{exact_current} can be combined and rewritten as 
\begin{multline}
\label{ddft2 planar}
\frac{1}{D_0} \, \frac{\partial \rho(z_1,t)}{\partial t} = \frac{\partial}{\partial z_1}  \Bigg( \frac{\partial \rho(z_1,t)}{\partial z_1}
+  \rho(z_1,t) \frac{\partial \beta V_{\text{ext}}(z_1)}{\partial z_1} \\
- 2\pi \int_{-\infty}^{\infty} dz_2 \, (z_1-z_2)\, \rho^{(2)}(z_1,z_2,r_2^{*},t) \Bigg),
\end{multline}
where $r_2^{*}\!=\!\sqrt{1-(z_1-z_2)^2}$ for the particle diameter set to unity. 
This corresponds to evaluating the two-body density only on 
the contact shell where the interparticle forces act. 
The force-DDFT generates the dynamics of the density profile in the virial realm, which contrasts and complements the standard potential-DDFT, which we recall in the following.

\subsection{Potential-DDFT}\label{potentialDDFT}

The current for potential-DDFT \cite{archer_evans} is given by
\begin{multline}
\label{approx_current}
\!\!\!\!\vec{j}(\vec{r},t) \!=\! -D_0 \, \rho(\vec{r},t)
\nabla_{\vec{r}}\Bigg( \ln(\rho(\vec{r},t))
+ \beta V_{\text{ext}}(\vec{r}) \\
-  c^{(1)}_{\text{p}}(\vec{r},t) \Bigg). 
\end{multline}
In the construction of the force-DDFT, in subsection \ref{forceDDFT}, we applied an adiabatic approximation to the two-body density, $\rhotwo$, 
and thus to the entire average interaction force. 
This approach explicitly implements the idea of instantaneous equilibration of $\rhotwo$ at each time-step.  
Here we exploit an equilibrium sum-rule (see Reference \cite{archer_evans}) to approximate the average interaction force using the gradient of the one-body direct correlation function, $c^{(1)}_{\text{p}}$, which results in equation 
\eqref{approx_current}.
The consequence of making this approximation is that the potential-DDFT operates within the 
compressibility realm. The long-time limit of the density time-evolution then reduces to that of the potential-DFT.
As already pointed out in the previous subsection, the current must vanish at equilibrium. In the present case this implies that the sum of terms in parentheses in \eqref{approx_current} must vanish, which recovers the (gradient of) the EL equation \eqref{EL potential}.

For the present case of planar geometry, combining equations \eqref{continuity} and \eqref{approx_current} yields 
the following one-dimensional equation of motion
\begin{multline}
\label{ddft1 planar}
\frac{1}{D_0} \, \frac{\partial \rho(z,t)}{\partial t} = \frac{\partial}{\partial z} \Bigg( \frac{\partial \rho(z,t)}{\partial z} + \rho(z,t) \frac{\partial \beta V_{\text{ext}}(z)}{\partial z} \\
- \rho(z,t) \frac{\partial c^{(1)}_{\text{p}}(z,t)}{\partial z} \Bigg)
\end{multline}
for the density profile. This can be compared with the exact equation \eqref{ddft2 planar}. 
Note that if we follow the adiabatic approximation scheme on the one-body density functional $\rhotwo[\rho]$, then we get back equation \eqref{ddft1 planar} but with $c^{(1)}_{\text{f}}$, defined by equation \eqref{def c1}, instead of $c^{(1)}_{\text{p}}$.

\subsection{Numerical results for hard-spheres in a harmonic-trap}\label{numericaltrap}

In order to compare the predictions of force-DDFT with those of potential-DDFT we consider a simple benchmark test of the relaxational dynamics. 
The density is first equilibrated to a planar harmonic external potential, $\beta V_{\text{ext}}(z)\!=\!A(z-z_0)^2$, where we use the values $A\!=\!0.75$ and $z_0\!=\!5$ inside of a computational domain covering the range from $z\!=\!0$ to $z\!=\!10$. 
At time $t\!=\!0$ we instantaneously switch the harmonic-trap amplitude to the value $A\!=\!0.5$ and then use either equation \eqref{ddft2 planar} or \eqref{ddft1 planar} to calculate the relaxational time-evolution of the density towards the equilibrium state of the new trap.   
The time integration of equations \eqref{ddft2 planar} and  \eqref{ddft1 planar} is performed using forward Euler integration, which amounts to approximating the partial time derivative according 
to the following finite difference expression
\begin{equation*}
\frac{\partial \rho(z,t)}{\partial t} \approx \frac{\rho(z,t+\Delta t)-\rho(z,t)}{\Delta t},
\end{equation*}
where $\Delta t$ is the time-step. 
In practice, the numerical realization of equilibrium as a long-time limit of the dynamics may be difficult to achieve due to the accumulation of discretization errors over many time-steps.

\begin{figure*}
\center{\includegraphics[width=0.95\linewidth]{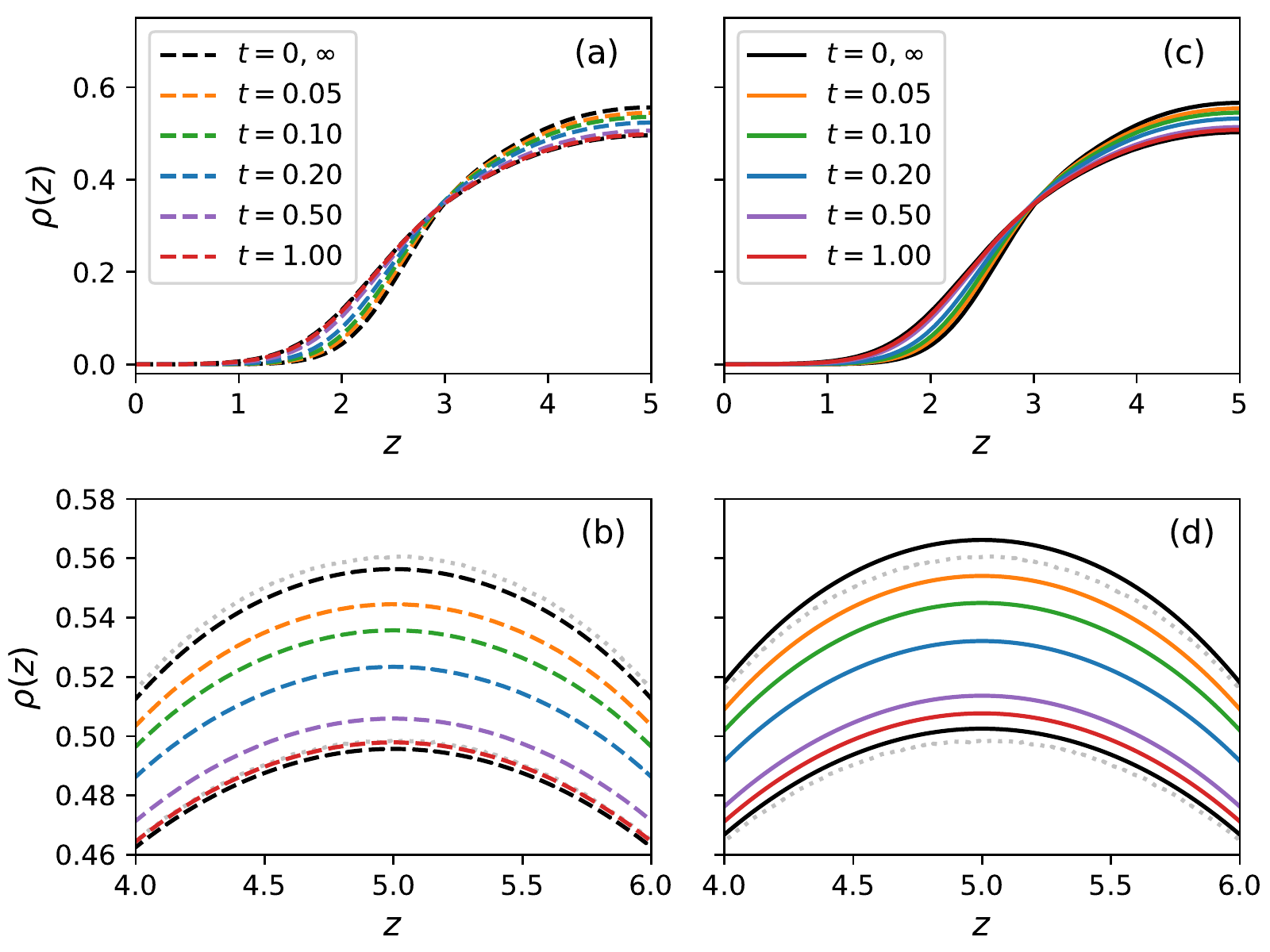}}
\caption{ {\bf Transient dynamics in a harmonic-trap.} Time-evolution
  of the density following a discontinuous change in the trap
  amplitude from $A\!=\!0.75$ to $0.5$ at time $t\!=\!0$.  The left
  panels (a) and (b) show the density obtained from
  potential-DDFT and the right panels (c) and (d) show the density
  from force-DDFT. The black lines
    (dashed for potential-DFT, solid for force-DFT)
    give the equilibrium initial and final states. 
    The silver dotted lines in the lower row, in panels (b) 
    and (d), show the initial equilibrium state and the final
    equilibrium state, as obtained from grand-canonical Monte-Carlo
    (GCMC) simulations.
\label{trap_fig}
}
\end{figure*}

The left panels (a) and (b) of Figure \ref{trap_fig} show the time-evolution of the density obtained 
from the potential-DDFT. In the upper panel we show only the left-half of the symmetric 
density profile and in the lower panel we show a zoom of the density peak. 
The black dashed curves indicate the equilibrium densities obtained from potential-DFT and 
can be compared with our grand-canonical Monte-Carlo simulation data \cite{frenkel}, given by the silver dotted lines.
The simulation is equilibrated for $10^5$ sweeps and sampled for $10^7$ sweeps, 
the box size is $30\times30\times20$, where the unit of length is a hard-sphere diameter, and on average 
there are $2142$ particles in the system. 
%

The colored dashed lines in Figure \ref{trap_fig} show density profiles obtained from potential-DDFT for a selection of 
different times, which we give in units of particle diameter squared over diffusion coefficient, $D_0$.
These results should be compared with those of the force-DFT and force-DDFT shown in the right panels (c) and (d)
of Figure \ref{trap_fig}, where we used the same colors as before to identify curves at equal times.
We clearly see that, in this case, the force-DDFT relaxes more slowly than potential-DDFT, which implies that the average (repulsive) interaction force is stronger in the latter approximation. 
A possible explanation for that phenomenon is that the hard-sphere system has a very harshly repulsive 
interparticle potential, which strongly influences the spatial distribution of the particles and which is 
captured more effectively in force-DDFT. 
{

The implementation of potential-DDFT is rather quick and simple since it only involves one-body functions and requires only a single Picard update at each time-step. On the other hand the force-DDFT is way more demanding since it involves solving the OZ equation at every time-step and then also requires a Picard update of the density. Not only are the analytical expressions more complicated, but also the numerical computational work. The shown curves therefore took significantly more computational time to be obtained, but they can nevertheless be calculated 
to high accuracy.

\section{Conclusions and outlook}\label{conclusions}
Starting from fundamental principles of Noether invariance, 
we have developed a force-based theory for the density profile both in- and out-of-equilibrium.
The equilibrium theory shows that density profiles can be calculated via the virial route by following our 
explicit force-DFT scheme. This situation can be contrasted with the standard potential-DFT that is known to follow the compressibility route.
The latter is a well-used result and often a crucial test in a significant number of DFT studies, so we provided a mathematical proof that explicitly shows that the planar hard-wall contact density 
from potential-DFT is given by the reduced \textit{compressibility} pressure.

Our analytical proofs have been tailored to highlight the different
outcomes from the two routes.  If we had access to the exact Helmholtz
free energy functional, then there would be no route-dependency.  A
more general proof of the contact theorem (shown in Appendix
\ref{general_contact} or in Reference \cite{hansen2013,lebowitz1960, lovett1991}) would then be sufficient.  We thus suggest to exploit the differences between the density profiles from the virial
and compressibility routes to test, scrutinize, and ultimately attempt
to improve, approximate DFT schemes.  Working with inhomogeneous
two-body correlation functions, as implemented explicitly in the
force-DFT, is both analytically and numerically more demanding than
using the standard potential-DFT scheme. However, facing the increase
in complexity is rewarded by gaining deeper insight into the
theoretical structure of DFT. Moreover working on the two-body level
allows to explicitly incorporate the pairwise interparticle
interactions and take direct account of their influence on the spatial
distribution of the particles. While carrying out 
force-DFT calculations comes at an increased numerical cost, the additional workload (both in terms of
   implementation and runtime) is far from prohibitive and 
   practical research can be efficiently performed.

  The distinction between the virial and
compressibility routes is known to be important in the
integral-equation theory of bulk liquids \cite{hansen2013}. Here we
reveal an analogous scenario for the theory of inhomogeneous 
fluids, which is an interesting result in its own right.
While both approaches, the conventional potential-DFT and the
   force-DFT, construct the density-functional dependencies in
   alternative forms, both approaches start from the same
   approximation for the excess free energy functional. This offers
   clear pathways towards improved theories that enforce
   self-consistency in a variety of ways. 
For example, the virial and compressibilty routes could be 
mixed in the spirit of liquid-state integral equation 
theories, using approximations analogous to the  
Rogers-Young \cite{RogersYoung} or Carnahan-Starling theories 
\cite{CarnahanStarling}. 
Another possibility would be to enforce the exact core-condition on the total correlation function $h(\vec{r}_1,\vec{r}_2)$ and this improve the description of the inhomogeneous two-body correlations.

A particularly appealing feature of the force-DFT is that it naturally
generalizes to treat nonequilibrium systems. At the most fundamental
level, particles are moved by forces, rather than by potentials, and
hence forces form a solid basis for developing a dynamical theory
\cite{delasheras2018velocityGradient,delasheras2020fourForces,schmidt2013pft,schmidt2021pft}.
The adiabatic approach that we have employed closes the dynamical
description on the level of the one-body density. On this basis we
have explored the dynamical behaviour of hard-spheres inside of a
harmonic-trap under a temporal switching protocol. We found that the
density dynamics that follow from potential- and force-DDFT differ
significantly from each other.  Not only are the equilibrium
(long-time) profiles different, but so are the relaxation rates.


The starting equations of force-DDFT, namely equations
\eqref{continuity} and \eqref{exact_current}, are exact.
If we would have access to the exact $\rhotwo$ as a functional of the one-body density, 
then we could calculate the exact time-evolution of $\rho$. 
As this information is
not available, we close the theory by making an adiabatic
approximation for $\rhotwo$, thus assuming that it equilibrates at each time-step. 
This thinking is also
captured in the adiabatic construction of power functional theory
\cite{schmidt2013pft,schmidt2021pft}.

A point of interest is to attempt to close at a higher level of the
correlation function hierarchy, with the aim to provide a
first-principles superadiabatic dynamical theory. It is hard to
conceive that such progress could made without a force-based approach.
The force-DFT that we present here thus represents a first step
towards full treatment of nonequilibrium.  Furthermore force-DDFT,
when compared with potential-DDFT, has the clear benefit that the
average interparticle interaction force does not appear automatically
as a gradient term in the exact equation \eqref{exact_current}.  If
this were the case, as it is in potential-DDFT, it would exclude
  \textit{de facto} all nonconservative forces. 
Force-DDFT thus
leaves the door open for future studies of driven systems 
such as systems with shear flows. It is well known that 
the adiabatic approximation within standard DDFT fails for shear fields 
\cite{aerov2014}. 
However, it would be interesting to investigate shear flows with higher order 
force-DDFT to check on the validity of these considerations and 
approximations.

Our derivation of the force balance (YBG) relationship from local
Noether invariance in section \ref{noether} is based on considering a
local displacement field $\eps(\rv)$. This object bears similarities
with the vector field that maps between positions in the Lagrangian
and Eulerian picture in continuum mechanics. The connections between
this thinking and a local density functional treatment were recently
explored by Sprik. Specifically he considered the case of dielectric
fluids \cite{sprik2021molPhys}.  Our more microscopic formulation
could possibly help to shed some light on the relationship of the
continuum mechanical force balance and the DFT equilibrium equation.
The connections to the crystalline state and crystal deformations are
also worth exploring, as addressed by Sprik within continuum mechanics
\cite{sprik2021jcp}, by Fuchs and coworkers from a more microscopic
point of view \cite{walz2010,haering2015} and recently by Lin et
al.\ \cite{lin2021} within DFT.

We have shown that the hard-sphere system is described within
   fundamental measure theory to a good level of self-consistency. 
   Going beyond hard-spheres and exploring the force-DFT for functionals 
   that describe interparticle attraction would be interesting. Such work 
   could be already revealing in the context of the standard mean-field 
   functional \cite{evans1992}.

In the context of power functional theory
\cite{schmidt2021pft} the force-based theories could play a role in the
description of the adiabatic state as applied to bulk and interfaces
of active Brownian particles \cite{krinninger2016,krinninger2019,
  hermann2019pre,hermann2019prl,hermann2021molPhys}, to flow phenomena
in overdamped systems \cite{stuhlmueller2018prl,
  delasheras2020fourForces}, to shear
\cite{treffenstaedt2020shear,jahreis2019shear} and the van Hove
function \cite{treffenstaedt2021,treffenstaedt2022}.

We would expect the treatment of long-ranged forces as they occur in charged 
 to require extra care in dealing with divergent integrals. Nevertheless, 
 application of force-DFT to Coulombic systems could be revealing for the 
 behaviour of the electrical double layer \cite{haertel2017review}, the differential capacitance 
   \cite{cats2021differentialCapacitance}, as well as for long-ranged decay of correlations, as 
   recently explored for the restricted primitive model \cite{cats2021decayLength}.

Burke and his collaborators have recently put forward a new approach
to electronic DFT. Their `blue electron approximation'
\cite{mccarty2020} offers a concrete way to work efficiently at finite
temperatures within what they call the conditional probability DFT
\cite{pederson2022}.  In the high-temperature limit an analogy to
Percus' classical test particle limit arises
\cite{mccarty2020}. As their method works on the two-body level
cross fertilization with our present approach is not inconceivable.

Molecular DFT generalizes classical DFT to systems with orientational
degrees of freedom, see e.g.\ References
\cite{teixera1991,groh1996dipolar,levesque2012jcp,jeanmairet2013jcp}. Various
ingeneous ways of dealing efficiently with the associated numerical
burdens of accounting for the molecular Euler angles have been
formulated, see e.g.\ Reference
\cite{ding2017sphericalHarmonics}. Whether the present approach can
help to describe the corresponding forces and torques in such systems
is an interesting point for future work. Also going beyond the planar
(effective one-dimensional) geometry and addressing fully
inhomogeneous three-dimensional situations
\cite{edelmann2016,tretyakov2016,stopper2017,stopper2018} constitutes
an exciting, yet formidable, research task.

As the two-body correlation functions upon which the force-DFT is
built are directly accessible via many-body simulation (see Reference
\cite{dijkstra2000dcf} aimed at the direct correlation function), one
can wonder whether using simulations data as input would allow to
construct force-DFT approximations. This could possibly be aided by
machine-learning techniques \cite{cats2021ml}.

The two-body density gives information about the probability to find a particle at position $\vec{r}_2$ given 
that there is a particle at position $\vec{r}_1$. 
This enables the pair interaction forces acting within the fluid 
to be analysed in detail. 
Moreover, multiplying the two-body density with the gradient of 
the pair-potential allows the average pair interaction force to 
be calculated explicitly and thus, 
in the case of hard interparticle interactions, incorporates the particle geometry directly.

\acknowledgments

SMT and JMB thank
G. T. Hamsler for her critical judgement and for taking the time to go through the whole manuscript 
several times.
MS acknowledges useful discussions with Daniel de las Heras.
This work is partially supported by the German Research Foundation (DFG) via Project No. 436306241.

\appendix

\section{Canonical transformation}\label{canonical}

The transformation given by
\eqref{EQcanonicalTransformationCoordinates} and
\eqref{EQcanonicalTransformationMomenta} is canonical and it hence 
preserves the phase-space volume element.  
That the transformation is canonical can be demonstrated by considering a generating function ${\cal G}$
\cite{goldstein2002}, which for the present transformation
has an explicit form given by
\begin{align}
  {\cal G} &= \sum_{i=1}^N \pv'_i\cdot (\rv_i +\eps(\rv_i)).
  \label{EQcanonicalTransformationGenerator}
\end{align}
As ${\cal G}$ is a function of the original coordinates and of the new
momenta, the transformation equations are generated via $\rv'_i \!=\!
\partial {\cal G}/\partial \pv'_i$ and $\pv_i \!=\! \partial {\cal
  G}/\partial \rv_i$.  Using the explicit form
\eqref{EQcanonicalTransformationGenerator} and expanding to lowest
order in $\eps(\rv)$ yields
\eqref{EQcanonicalTransformationCoordinates} and
\eqref{EQcanonicalTransformationMomenta} in a straightforward way.

The canonical generator $\cal G$ defined in equation
\eqref{EQcanonicalTransformationMomenta} is a function of the original
coordinates $\rv_1,\ldots,\rv_N$ and of the new momenta
$\pv_1',\ldots,\pv_N'$. For the case of such dependence the original
Hamiltonian $H$ and the transformed Hamiltonian $H'$ are related by
the general transformation \cite{goldstein2002}:
\begin{align}
  H' = H+\frac{\partial \cal G}{\partial t}.
  \label{EQnewHamiltonian}
\end{align}
As the generator \eqref{EQcanonicalTransformationMomenta} carries no
explicit time dependence, the last term in equation
\eqref{EQnewHamiltonian} vanishes, and $H'=H$. This invariance of the
Hamiltonian under the considered transformation implies the trivial
replacement of variables, i.e.\ that the transformed Hamiltonian
depends on the transformed coordinates and momenta,
i.e.\ $H'(\rv_1',\ldots,\rv_N',\pv_1',\ldots,\pv_N')$. Then by
construction, the equations of motion, when expressed in the new phase
space variables, are generated from the standard Hamiltonian
procedure: $ d\pv_i'/dt = -\partial H'/\partial \rv_i'$ and $
d\rv_i'/dt = \partial H'/\partial \pv_i'$.

\section{General derivation of the contact theorem}\label{general_contact}

The following appendix shows a derivation of the contact theorem appropriate 
to situations in which all quantities are known exactly. For this reason we use the 
generic notation $c^{(1)}$ and $P$ for the one-body direct correlation function and the pressure, respectively.

Let us consider a hard-wall such that the distance of closest approach of a particle is located at $z\!=\!0$.
We assume that the system reaches a bulk-like state at and around a (large) distance $L$
away from the wall. In order to have a closed system in the
$z$-direction, we consider a second `ultrasoft' wall that vanishes
for $z\!<\!L$, and then gives a slowly rising energy penalty upon increasing $z$, which ultimately diverges 
$V_{\text{ext}}(z\to\infty)\!=\!\infty$.

We recall the global Noether identity of vanishing total interparticle
force
\begin{align}
  \int_{-\infty}^\infty dz \, \rho(z) \frac{dc^{(1)}(z)}{dz} &= 0,
  \label{EQhardWallNoetherVanishingForce}
\end{align}
where $c^{(1)}(\rv)=-\beta \delta F_{\text{exc}}[\rho]/\delta\rho(\rv)$ is
the one-body direct correlation function. The integrand in
\eqref{EQhardWallNoetherVanishingForce} is, up to a factor of thermal
energy, the locally resolved interparticle force density,
$k_BT\rho(\rv)\nabla_{\rv} c^{(1)}(\rv)$, acting in the $z$-direction. One can
argue equivalently and independently (see,
  e.g.\ \cite{hermann2021noether}), that
\eqref{EQhardWallNoetherVanishingForce} holds on the basis of
Newton's third law.

Here we rather start from the alternative form
\begin{align}
  \int_{-\infty}^\infty dz \,  c^{(1)}(z) \frac{d\rho(z)}{dz}=0,
  \label{EQhardWallNoetherVanishingForceByParts}
\end{align}
which is straightforwardly obtained from the Noether sum-rule
\eqref{EQhardWallNoetherVanishingForce} via integration by parts;
circumstances must be such that the boundary terms at infinity vanish.
More significantly, within a DFT context, it is the form
\eqref{EQhardWallNoetherVanishingForce} that is the primary result
from applying Noether's theorem to the invariance of the excess free
energy functional $F_{\text{exc}}[\rho]$ upon spatial shifting of the
system \cite{hermann2021noether}.

Here we proceed directly with the form
\eqref{EQhardWallNoetherVanishingForceByParts}, treating three spatial
regions separately: the vicinity of the hard-wall, $-\Delta\!<\!z\!<\!\Delta$,
where $\Delta$ is a small parameter (as compared to all other
lengthscales in the sytem); the region from the wall to the bulk-like
state, i.e.\ $\Delta\!<\!z\!<\!L$; and the soft wall region, $z\!>\!L$. In the
following, the limit $\Delta\!\to\! 0$ is implicit.

In the vicinity of the hard-wall we can identify the leading term as
\begin{align}
  \int_{-\Delta}^\Delta dz \, c^{(1)}(z) \frac{d\rho(z)}{dz}
  &= \int_{-\Delta}^\Delta dz \, c^{(1)}(z)\delta(z)\rho(0) \notag\\
  &=c^{(1)}(0)\rho_{\text{w}}
  \notag\\&
  =\rho_{\text{w}}\ln(\rho_{\text{w}}) - \beta\mu\rho_{\text{w}},
  \label{EQhardWallNoetherRegionOne}
\end{align}
where $\rho_{\text{w}}\!=\!\rho(0)$ and in the last step we have used the
EL equation
\begin{align}
  c^{(1)}(z) &= \ln(\rho(z)) + \beta V_{\text{ext}}(z) - \beta \mu,
  \notag 
\end{align}
to express the one-body direct correlation function at contact, $c^{(1)}(0)$;
note that the external potential term gives no contribution as $V_{\text{ext}}(0^+)\!=\!0$; furthermore $c^{(1)}(z)$ is continuous at $z\!=\!0$.

In the region from between outside the wall and the bulk, i.e.\ for
$\Delta\!<\!z\!<\!L$, the external potential vanishes and we have
\begin{align}
  &\int_\Delta^L dz \, c^{(1)}(z)\frac{d\rho(z)}{dz} =
  \int_\Delta^L dz \, [\ln(\rho(z))-\beta\mu]\frac{d\rho(z)}{dz}\notag\\
  &\quad=[\rho(\ln(\rho)-1)-\beta\mu\rho]_{\rho_{\text{w}}}^{\rho_{\text{b}}}\notag\\
  &\quad= \rho_{\text{b}}(\ln(\rho_{\text{b}})-1-\beta\mu)
  - \rho_{\text{w}}(\ln(\rho_{\text{w}})-1-\beta\mu),
  \label{EQhardWallNoetherRegionTwo}
\end{align}
where in the first step we have again used the EL equation
\eqref{EQhardWallNoetherVanishingForce} and the bulk density is
defined as $\rho_{\text{b}}\!=\!\rho(L)$.

In the soft wall regime, i.e.\ for $L<z$, the density inhomogeneity is
so weak that a local density approximation becomes accurate and hence
\begin{align}
  \int_L^\infty \!\! dz \, c^{(1)}(z)\frac{d\rho(z)}{dz} &=
  \int_{\rho_{\text{b}}}^0 d\rho \, c^{(1)}(\rho)
  =f_{\text{exc}}(\rho_{\text{b}})\notag\\
  &  = -P - \rho_{\text{b}}(\ln(\rho_{\text{b}})-1)+\mu\rho_{\text{b}}.
  \label{EQhardWallNoetherRegionThree}
\end{align}
The upper limit in the density integral is $\rho(z\!\to\!\infty)\!=\!0$, and
$f_{\text{exc}}(\rho_{\text{b}})$ is the bulk excess free energy density per
volume as a function of $\rho_{\text{b}}$. The value at the upper boundary of
the density integration vanishes, as the system is infinitely dilute. 
The last step identifies the pressure $P$.

Adding up the three contributions \eqref{EQhardWallNoetherRegionOne},
\eqref{EQhardWallNoetherRegionTwo} and
\eqref{EQhardWallNoetherRegionThree} gives according to Noether
invariance \eqref{EQhardWallNoetherVanishingForceByParts} a vanishing
result. Rewriting yields
\begin{align}
  \rho_{\text{w}} = \beta P,
\end{align}
which is the general form of the hard-wall sum-rule.

\section{Planar hard-sphere force integral}
\label{Appendix_cOneForce}
In the following we derive the one-body direct correlation function
$c_\text{f}^{(1)}$, given by equation \eqref{eq:cf_hardsphere}, for hard-spheres in planar
geometry.  We start with the gradient of 
$c_\text{f}^{(1)}$, from equation \eqref{def c1}, namely
\begin{align}
\nabla_{\rv_1} c_{\text f}^{(1)}(\rv_1) = -  \int d \rv_2 \frac{\rho^{(2)}(\rv_1, \rv_2)}{\rho(\rv_1)} \nabla_{\rv_1} \beta\phi_{12}, \label{eq:A}
\end{align}
and as a first step exploit the planar geometry. The symmetry
simplifies the dependence on the position variables such that the one-body
distributions only depend on the $z$ coordinate.  The two-body density,
$\rho^{(2)}$, depends on  $z_1$, $z_2$ and $r_2$ (see Reference \cite{tschopp2021}).
The distance between the two particle positions is then $r_{12}\!=\!\sqrt{r_2^2+(z_1-z_2)^2}$.

As $c_\text{f}^{(1)}$ only depends on $z$, the gradient on the left
hand side of equation \eqref{eq:A} reduces to $\unit_z \, d/dz $, where
$\unit_z$ is the unit vector in the $z$-direction.  The
interparticle interaction potential, $\phi$, depends only on $r_{12}$. This allows us to rewrite the
gradient of $\phi$ as a derivative with respect to this distance,
$\unit_{r_{12}} \, d/{d r_{12}} $, where $\unit_{r_{12}}\!=\!
(\rv_1-\rv_2)/r_{12}$ indicates the radial unit vector.
Equation \eqref{eq:A} thus simplifies to
\begin{align}
\frac{d \, c_{\text f}^{(1)}(z_1)}{d z_1} \, \unit_{z} = - \int d \rv_2 \frac{\rho^{(2)}(z_1,z_2,r_2)}{\rho(z_1)} \frac{d \, \beta\phi_{12}}{d r_{12}} \, \unit_{r_{12}}. \label{eq:B}
\end{align}
We next express the $\rv_2$-integral in cylindrical
coordinates, such that the $z$-component of equation \eqref{eq:B}
becomes
\begin{align}
&\frac{d \, c_{\text f}^{(1)}(z_1)}{d z_1}  \label{eq:nablaC} \\
&= - \frac{2\pi}{\rho(z_1)} \! \int\limits_{-\infty}^\infty \!d z_2 \int\limits_0^\infty \! d {r}_2  \, r_2 \, \rho^{(2)}(z_1,z_2,r_2) \frac{d \, \beta\phi_{12}}{d r_{12}} \frac{(z_1-z_2)}{r_{12}}. \nonumber
\end{align}

To deal with the hard-sphere potential, $\phi$, we proceed as previously in equation \eqref{eq:ext}. 
We therefore multiply the integrand in
equation \eqref{eq:nablaC} by $1\!=\! {\rm e}^{\beta \phi_{12}} {\rm e}^{-\beta
  \phi_{12}}$.  The second Boltzmann factor can be grouped together
with the derivative of the interaction potential as ${\rm e}^{-\beta
  \phi_{12}} \frac{d \beta\phi_{12}}{d r_{12}} \!=\! - d {\rm e}^{-\beta
  \phi_{12}}/d r_{12}$.  For the
hard-sphere interaction potential the Boltzmann factor can be
identified as a step function, ${\rm e}^{-\beta \phi_{12}} \!=\!
\Theta(r_{12}-1)$, where $\Theta$ indicates the Heaviside step
function. The radial derivative then gives a Dirac delta
distribution,
\begin{align*}
\frac{d \, \Theta(r_{12}-1)}{d r_{12}} =\delta(r_{12} - 1)  =  \frac{ \delta(r_2 - r_2^*)}{ \left| r_{2}/r_{12}\right|}, 
\end{align*}
where $r_2^*\!=\! \sqrt{1-(z_1-z_2)^2}$ is the cylindrical radial distance
at contact for given coordinates $z_1$ and $z_2$.
This yields
\begin{align*}
\frac{d}{d z_1} c_{\text f}^{(1)}(z_1) = - \frac{2 \pi}{\rho(z_1)} \int\limits_{-\infty}^\infty d z_2 \, (z_1-z_2) \rho^{(2)}(z_1,z_2,r_2^*).
\end{align*}
To obtain the the desired equation \eqref{eq:cf_hardsphere}, we then integrate with respect to $z_1$ from 
$0$ to $z$.

\end{document}